\documentclass[a4paper,11pt]{article}
\pdfoutput=1 

\usepackage{jcappub} 

\usepackage[T1]{fontenc} 

\usepackage{graphicx,natbib,amssymb,amsmath,stmaryrd,makecell}
\usepackage{txfonts}
\usepackage{xcolor}
\usepackage{hyperref}
\usepackage{xspace,dsfont}
\usepackage{caption, lipsum, comment}

\graphicspath{{Figures/}}

\newcommand{\keV}{{\rm keV}}

\newcommand{\expf}[1]{{{\rm e}^{#1}}}

\newcommand{\vgh}{{\hat{\boldsymbol\gamma}}}

\newcommand{\vb}{{\boldsymbol{\beta}}}
\newcommand{\vbh}{{\boldsymbol{\hat{\beta}}}}

\newcommand{\id}{{\,\rm d}}

\newcommand{\beq}{\begin{equation}}   %

\newcommand{\eeq}{\end{equation}}   %

\newcommand{\beqa}{\begin{eqnarray}}   %

\newcommand{\eeqa}{\end{eqnarray}}   %

\newcommand{\bealf}[1]{\begin{align} #1 \end{align}}

\newcommand{\beal}{\begin{align}}
\newcommand{\enal}{\end{align}}

\newcommand{\bspl}{\begin{split}}

\newcommand{\espl}{\end{split}}

\newcommand{\bsub}{\begin{subequations}}

\newcommand{\esub}{\end{subequations}}

\newcommand{\bmulti}{\begin{multline}}   %

\newcommand{\beqm}{\begin{mathletters}}   %

\newcommand{\eeqm}{\end{mathletters}}   %

\newcommand{\me}{m_{\rm e}}

\newcommand{\Te}{T_{\rm e}}

\newcommand{\The}{\theta_{\rm e}}

\newcommand{\sigT}{\sigma_{\rm T}}

\newcommand{\vek} [1]{\mbox{\boldmath${#1}$\unboldmath}}

\newcommand{\sterling}[2]{\genfrac{[}{]}{0pt}{0}{#1}{#2}}

\newcommand{\oDnu}{{\mathcal{\hat{D}}_{\nu}}}

\newcommand{\oOnu}{{\mathcal{\hat{O}}_{\nu}}}

\usepackage{comment}

\newcommand{\bra}[1]{\langle #1 \vert}
\newcommand{\ket}[1]{\vert #1 \rangle}

\newcommand{\Dbo}[3]{{{^{#1}}\mathcal{\hat{D}}^{#3}_{#2}}}

\usepackage{bbold}
\usepackage{bbm}
\def\i{\mathbbm{i}}

\title{Derivation of the Kompaneets equation using the boost operator approach}

\author[a]{Alex Hoey}
\author[a]{Jacob Long}
\author[a]{Jens Chluba}

\emailAdd{alex.hoey@student.manchester.ac.uk}
\emailAdd{jacob.long@student.manchester.ac.uk}

\affiliation[a]{Jodrell Bank Centre for Astrophysics, School of Physics and Astronomy, The University of Manchester, Oxford Road, Manchester, M13 9PL, U.K.}

\date{Feb 2026}

\begin{document}

\abstract{The repeated scattering of photons by thermal electrons at low temperatures is described by the Kompaneets equation and its generalized forms that include anisotropies and higher order temperature corrections. In this work, we use the boost operator approach to derive the related expressions in a transparent way that showcases the generality of the formalism and its application to radiative transfer problems. We consider the simplest form of the Kompaneets equation for the scattering in isotropic media at the leading order in the electron temperature and then include anisotropies in the photon field, reproducing previously obtained expressions for the evolution equations. For this we use expressions for the scattering operator in the electron rest frame up to first order in the electron recoil, $\mathcal{O}(h\nu/\me c^2)$, but then work at all orders in the electron momentum, $p$, as easily obtained with the boost operator approach. This shows how specific transformation rules can be formulated that allow simplification of the otherwise cumbersome and repetitive calculations. 
%
%
We also confirm the expressions for higher order temperature corrections in isotropic media, highlighting the validity of the approach presented here. As part of the derivation, we find expressions for the boost operator in general boost directions which we believe will also be useful in other applications of the formalism.}

\maketitle
\flushbottom
\newpage

\section{Introduction}
\label{introduction}
In thermal plasmas, one of the most important interactions is due to the repeated Compton scattering between free electrons and photons. At low electron temperatures ($k \Te \lesssim 1\,\keV$), this process is described by the Kompaneets equation \citep{Kompa56} and its generalized forms to also include anisotropies in the photon field \citep{Bartolo2006, Pitrou2009, Chluba2012}.
Although the Kompaneets equation was originally developed to model the scattering of neutrons in nuclear reactions, it has found many applications in astrophysical plasmas and cosmology \citep{Zeldovich1969, Sunyaev1970mu, Blumenthal1970}. Most importantly, it is one of the key ingredients for the thermalization of primordial spectral distortions of the cosmic microwave background (CMB) \citep{Sunyaev1970SPEC, Burigana1991, Hu1993, Chluba2011therm} and, in its more general form, the modeling of the Sunyaev-Zeldovich (SZ) effect \citep{Zeldovich1969, Sazonov1998, Itoh98, Challinor1998, Chluba2012SZpack}.

The Kompaneets equation can be deduced in a number of ways. For isotropic media, it takes the simple form\footnote{All the physical constants take their common meaning unless stated otherwise.}
\begin{align}
\label{Kompaneets eq}
\frac{\text{d}n}{\text{d}\tau} =
\frac{k\Te}{\me c^2} \, \frac{1}{x^2}\,\frac{\partial}{\partial x} \left\{ x^4 \left[ \frac{\partial n}{\partial x} + n(1+n)\right]\right\},
\end{align}
where $n(\nu)$ is the occupation number of the isotropic photon field, $x=h\nu/k\Te$, is the dimensionless photon frequency, and $\tau = \int \sigT n_e c \id t$ is the Thomson scattering optical depth caused by free electrons with number density $n_e$. The first group of terms in Eq.~\eqref{Kompaneets eq} accounts for the energy transfer by Doppler diffusion, while the terms $\propto n(1+n)$ capture electron recoil and stimulated recoil. 

The basic steps for obtaining the Kompaneets equation involve considering the Boltzmann collision term for the Compton process, with an expression for the Compton scattering cross section for a thermal distribution of free electrons, and then performing a Fokker-Planck approximation in terms of the energy exchange between the photon and scattering electron \citep[see][for recent overview]{Oliveira2021Komp}. In the derivations, several integrals over the scattering angles of photons and electrons have to be performed. At higher orders in the energy exchange or when including anisotropies in the photons field, these integrals become quite cumbersome and extremely repetitive, begging the question if one cannot find a more transparent and systematic approach to performing the calculations, while focusing on the individual physical effects. 

In the rest frame of the moving electron, the general scattering cross section takes a much simpler form \citep[e.g.,][]{Jauch1976}. By performing Lorentz transformations of the photon field into this frame, applying the scattering operator, and then transforming back to the lab frame, we can directly describe the single momentum scattering process. Upon averaging over the thermal distribution of electrons, this yields the required result, an approach that has indeed been partially used in the past in the context of the SZ effect \citep[e.g.,][]{Sazonov1998}, with many explicit integrals required to simplify the intermediate results.

\begin{figure}
    \centering
    \includegraphics[width=1\linewidth]{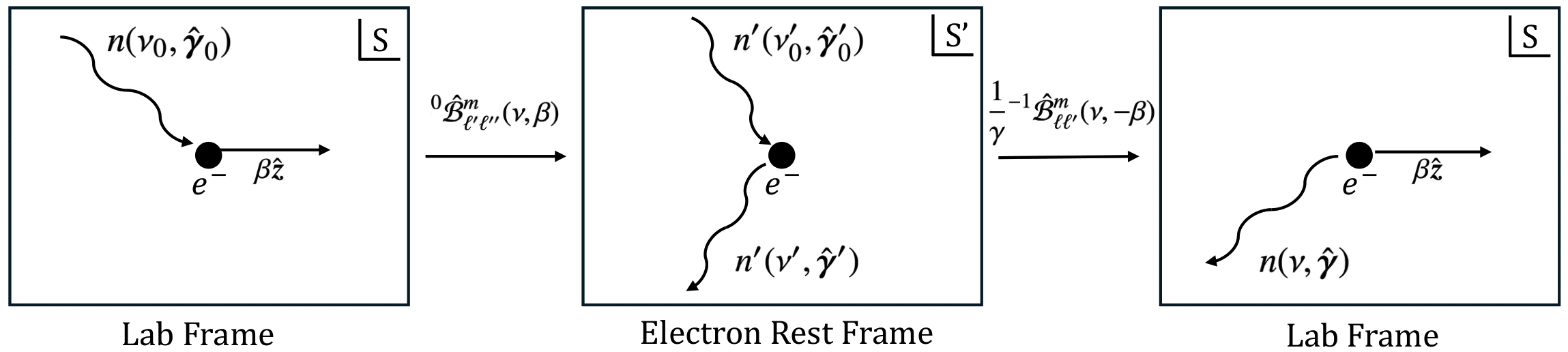}
    \caption{A schematic diagram of the boost operator approach. The lab frame occupation number, $n(\nu_0, \vgh_0)$, is boosted into the electron rest frame along the $z$-axis using the boost operator ${}^{0}\hat{\mathcal{B}}^{m}_{\ell' \ell''}(\nu,\beta)$ where the scattering is evaluated. The scattered occupation number, $n'(\nu', \vgh')$, is then boosted back into the lab frame using the combination ${}^{-1}\hat{\mathcal{B}}^{m}_{\ell \ell'}(\nu,-\beta)/\gamma$, where the Doppler weight $-1$ and the $1/\gamma$ factor arise from the Lorentz transformation of the scattering optical depth (see section \ref{sec:Lab frame col term}).}
    \label{fig:schematic}
\end{figure}
In this work, we use the recently developed boost operator approach \citep{ChlubaBO25} to derive the Kompaneets equation with generalizations. The elegance of our derivation is that most of the integrals can be avoided as the boost operator directly provides the required mapping between frames (see section~\ref{sec:BO}). A general schematic for our approach is shown in Fig.~\ref{fig:schematic}. 
The advantage of our method is that in the electron's rest frame the process becomes much easier to describe. Furthermore, we are able to obtain an exact, all-orders in the electron momentum, $p$, expression for the scattering operator at varying order in the electron recoil effect, $\propto h\nu/\me c^2$. In this way, the physical origin of each term becomes transparent, and all that is required is expansions of final operators to orders in $p$, which can simply be done using the symmetry and recursion expressions given in \cite{Dai2014} and \cite{ChlubaBO25}. 
Expressions for these expansions can be obtained in powers of the energy shift generator $\oOnu = -\nu\partial_{\nu}$, eliminating the need for repeatedly carrying out complicated integrals. The thermal moments of $p^k$ can also be easily evaluated analytically \cite{CSpack2019}, leaving our expression as a function of the dimensionless $\The=k\Te/\me c^2$. The same approach has already been applied to the relativistic SZ effect for both the polarised \citep{Rosenberg2025prSZ} and unpolarised \citep{Chluba2026SZ} cases; however, here we also have to account for the electron recoil effect, which can be entirely neglected for the SZ effect, thus generalizing the previous results.

The paper is structured as follows. We first briefly introduce the aberration kernel \citep{Dai2014} and the boost operator \citep{ChlubaBO25} in sections~\ref{sub:Intro AK} and \ref{sec:BO}, respectively. Both the boost operator and aberration kernel should be seen as new 'special functions' describing the transformation of fields that can be expressed on a sphere, with unique properties and relations simplifying all the mathematical steps.
Importantly, in these sections we generalize the expressions to allow for arbitrary direction of the boost (section~\ref{Arbitrary boost direction}), an ingredient that is important for the final application.
In section~\ref{sec: Rest frame col term} we give the required Compton collision term in the electron rest frame before transforming it to the lab frame in section \ref{sec:Lab frame col term}. The Kompaneets equation is derived from our general expression at first order in the electron recoil in section \ref{Kompaneets Equation}. Section~\ref{sec: Anisotropies} provides a calculation of the anisotropic scattering corrections and confirms the results given by \cite{Chluba2012}. Finally, in section~\ref{app:2nd order temp} we consider second order temperature corrections and cross check with the work of \cite{Sazonov2000} to confirm the overall validity of our method.

\section{The aberration kernel}
\label{sub:Intro AK}
\noindent
In this section, we briefly introduce what will be a crucial tool in our derivation of the Kompaneets equation: The aberration kernel. In simplest terms, the kernel describes how the multipole coefficients of {\it frequency-independent} quantities that are described on a sphere transform between two frames in relative motion with one another. Consider some quantity $X$, which has spin-weighted harmonic coefficients ${}_s X_{\ell m}$ in the lab (observer) frame, $S$, and ${}_s X_{\ell m}'$ in the moving (electron) frame, $S'$. Assuming that the direction of the motion is aligned with the $z$-axis, and $\beta=\varv/c$ is the speed in units of the speed of light, these coefficients are determined by 
\bealf{
\label{app:AK}
{}_s X_{\ell m}'
&= 
\sum_{\ell'} {}^{d}_{-s}\mathcal{K}^{m}_{\ell \ell'}(\beta)\,{}_s X_{\ell'm},
}
where ${}^{d}_{s}\mathcal{K}^{m}_{\ell \ell'}(\beta)$ is the aberration kernel \citep{Challinor2002, Chluba2011ab, Dai2014}.
The quantity $s$ denotes the spin weight of the coefficient and $d$ is the Doppler weight which depends on how the quantity Lorentz transforms. Quantities such as photon occupation number, which we will use below, have Doppler weight $d=0$ whilst temperature fields have a Doppler weight of $d=1$.\footnote{An observable of general Doppler weight, $d$, will transform as $X'(\boldsymbol{\hat{n}}')
= 
\left(\nu'/\nu\right)^d\,
X(\boldsymbol{\hat{n}}\,[\boldsymbol{\hat{n}}']),
$ where $\nu$ and $\nu'$ are the photon frequencies in the lab and moving frame, respectively, and $\boldsymbol{\hat{n}}$, $\boldsymbol{\hat{n}}'$ are the directions on the sky in each frame.} 

\begin{figure}[hbt!]
    \centering
    \includegraphics[width=0.85\linewidth]{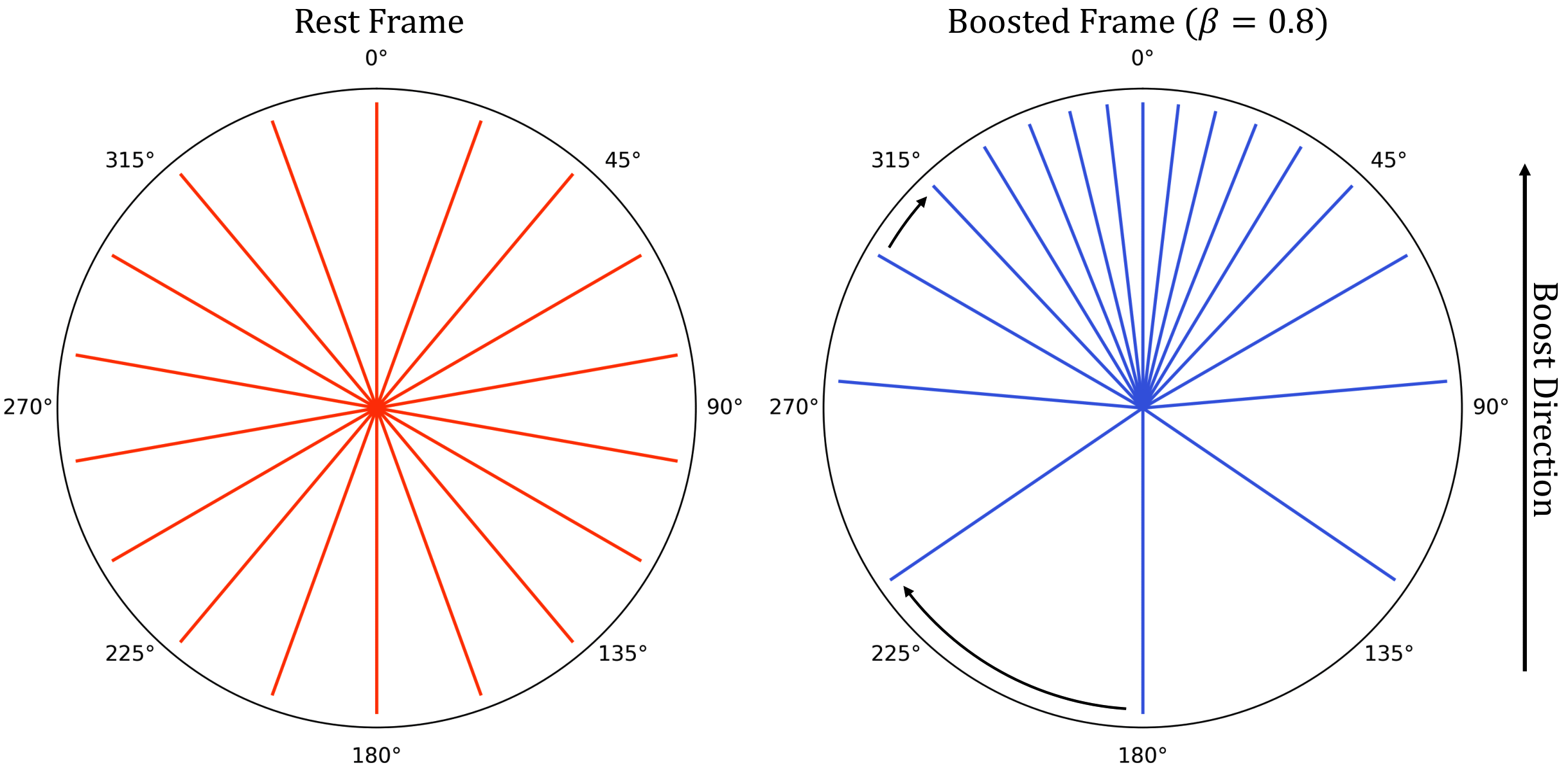}
    \caption{Representations of a distribution as seen from two frames boosted relative to each other. The rest frame distribution (left) is isotropic and becomes distorted towards the direction of the boost when viewed from the boosted frame (right). The magnitude of the boost shown here is $\beta=0.8$.}
    \label{fig:AK}
\end{figure}

Due to the relative motion of the two frames, to an observer in $S'$ the apparent arrival directions of CMB photons are deflected slightly. The change in angle can be approximated as $\Delta \theta \sim \beta$ and leads to a redistribution of power between the multipoles, $\ell$. 
A two-dimensional representation of this is shown in Fig.~\ref{fig:AK} where an apparent isotropy in the rest frame becomes anisotropic in a boosted frame. We can imagine these depictions as projections onto the sky giving us azimuthal symmetry in both frames about the boost axis.
Mixing between neighboring modes $(\ell \pm 1)$ is most likely with weaker contributions from further neighboring modes $(\ell \pm 2,\ell\pm 3, \ldots)$ \citep{Chluba2011ab}.
Since the Kompaneets equation describes the evolution of CMB photons due to Compton scattering with electrons, understanding how the multipole coefficients we observe relate to those in the electron rest frame is essential for our method. The aberration kernel provides and exact function which maps between the two frames. Explicitly, the aberration kernel is defined as 
\bealf{
\label{eq:Aberration Kernel}
{}^{d}_{s}\mathcal{K}_{\ell' \ell}^{m}(\beta)
&= 
\int \text{d}^2\boldsymbol{\hat{n}}'\,
\frac{{_{-s}}Y^{*}_{\ell' m}(\boldsymbol{\hat{n}}')\,{_{-s}}Y_{\ell m}(\boldsymbol{\hat{n}})}{\left[\gamma \, (1-\boldsymbol{\beta} \cdot \boldsymbol{\hat{n}}')\right]^{\,d}},
}
where $\boldsymbol{\hat{n}}=\boldsymbol{\hat{n}}[\boldsymbol{\hat{n}'}]$ and $\boldsymbol{\hat{n}'}$ are the directions on the sky in the respective frames and $\gamma$ is the usual Lorentz factor. Note we have defined the \textit{observer} kernel, which fixes the sign convention for $\boldsymbol{\beta}$ such that photons received in the direction of the motion are deflected towards the direction of the motion. 

As highlighted in \citep{Dai2014}, one can also think of the problem in Hilbert space and use the boost generator along the $z$-axis, $$\hat{Y}_z = - \i\left(\mu + \sqrt{1-\mu^2} \partial_\phi\right),$$ with $\boldsymbol{\hat{\beta}} \cdot \boldsymbol{\hat{n}} = \mu=\cos\theta$ and azimuthal angle $\phi$, to generate the kernel expressions. This leads to the more intuitive bra-ket notation of the kernel \citep{Dai2014}
\bealf{
\label{app:bra_ket}
^{d}_{s}\mathcal{K}_{\ell' \ell}^{m}(\beta)
&= 
\bra{s {\ell'} m} e^{\i\eta \hat{Y}_z} \ket{s \ell m},
}
where $\eta=\tanh^{-1}(\beta)$ is the rapidity. In Appendix \ref{app:Kernel symmetries}, we give a brief summary of some useful kernel symmetries and recurrence relations as given by \cite{Dai2014}. A more complete discussion of these is given in section~\ref{sub:BO symmetries}, once we have introduced the boost operator.

\subsection{Generalising to arbitrary boost direction}
\label{Arbitrary boost direction}
We now generalize the expression for the aberration kernel to arbitrary boost direction using the operator language of \cite{Dai2014}. The expression for the case in which $\vbh$ is collinear with $\boldsymbol{\hat{z}}$ is given above by Eq.~\eqref{app:bra_ket}. Since the kernel elements in this coordinate frame are straightforward to calculate, we perform a rotation into this frame from our original coordinate frame in which the boost direction is completely arbitrary. We define our rotation such that
\bealf{
\label{app:rotation}
R\boldsymbol{\hat{z}} = \vbh,
}
where $R$ is the appropriate rotation matrix. In addition, we define $\hat{U} (R)$ as the unitary operator that implements this passive rotation in Hilbert space. Now considering the dot product between the vector boost generator $\boldsymbol{\hat{Y}}$ and $\vbh$, we find
\bealf{
\label{app:dot product}
\vbh \cdot \boldsymbol{\hat{Y}} = \hat{\beta}_i \hat{Y}_i = \hat{\beta}_i\, \left[\hat{U} (R)\,\hat{U}^{\dagger} (R)\right]\,
\hat{Y}_i\,
\left[\hat{U} (R)\,\hat{U}^{\dagger} (R)\right]
= \hat{U} (R)\,
\hat{\beta}_i\,
\left[\hat{U}^{\dagger} (R)
\hat{Y}_i
\hat{U} (R)\right]\,
\hat{U}^{\dagger} (R).
}
A standard similarity transform for an operator $\hat{O}_i$ is given by $\hat{O}_i'=\hat{U}(R)\,\hat{O}_i\,\hat{U}^{\dagger} (R)$. Thus we can identify that $\hat{U}^{\dagger}(R)\,\hat{Y}_i\,\hat{U} (R)$ is the transform for the inverse rotation $[R_{ij}]^{-1}$, a product of us defining $R$ as the passive rotation. Referring to Eq. (\ref{app:rotation}) we have that $\hat{\beta}_i\,[R_{ij}]^{-1}=\hat{z}_j$ and thus,
\bealf{
\label{app:dot product 2}
\vbh \cdot \boldsymbol{\hat{Y}} = \hat{\beta}_i \hat{Y}_i = \hat{U} (R)\,
\hat{\beta}_i\,
[R_{ij}]^{-1}\,
\hat{Y}_j\,
\hat{U}^{\dagger} (R)
= \hat{U} (R)\,
\hat{z}_j\,
\hat{Y}_j\,
\hat{U}^{\dagger} (R).
}
Thus, the boost along $\vbh$ is related to the boost along the $z$-axis by
\bealf{
\label{app:similarity_transform}
e^{i\eta \,\vbh \cdot \boldsymbol{\hat{Y}}}
= e^{i\eta\,\hat{U} (R)\,
\hat{Y}_z\,
\hat{U}^{\dagger} (R)}
= \hat{U} (R) \, \expf{i\eta {\hat{Y}_z}}\,\hat{U}^\dagger (R),
}
where we can perform the step in the final equality since $\hat{U} (R)$ is unitary. The kernel elements for an arbitrary boost direction are then given by
\bealf{
\label{app:arbitrary_kernel}
^{d}_{s}\mathcal{K}_{\ell' \ell}^{m'm}(\boldsymbol{\beta})
= \bra{s\ell'm'}\,\hat{U} (R) \, \expf{i\eta {\hat{Y}_z}}\, \hat{U}^\dagger (R)\,\ket{s\ell m}.
}
We proceed by inserting resolutions of the identity. Note that these do not cause mixing of $\ell$ states. We thus have
\bealf{
\label{app:identity_res}
^{d}_{s}\mathcal{K}_{\ell' \ell}^{m'm}(\boldsymbol{\beta})
= \sum_{m_1, m_2} \bra{s\ell'm'}\,\hat{U} (R) \,\ket{s \ell'm_1}\bra{s \ell'm_1}\, \expf{i\eta {\hat{Y}_z}}\,\ket{s \ell m_2}\bra{s \ell m_2}\,\hat{U} (R)^\dagger\,\ket{s\ell m}.
}
We recognize the rotation operator matrix elements as Wigner D-functions \citep[e.g.,][]{Varshalovich1988},
\bsub
\bealf{
\label{app:Wigner-D}
\bra{s \ell' m'}\,\hat{U} (R)\,\ket{s\ell' m_1}&\equiv 
D^{\ell'}_{m'm_1}(\phi,\theta,\psi),
\\
\bra{s \ell m_2}\,\hat{U}^\dagger (R)\,\ket{s\ell m}&
=\bra{s \ell m}\,\hat{U} (R)\,\ket{s\ell m_2}^*\equiv 
D^{\ell*}_{m m_2}(\phi,\theta,\psi),
}
\esub
where $\psi$ is the roll angle that drops out of the final equation. This is intuitive; since we rotate to be collinear with $\boldsymbol{\hat{z}}$, rotation about this axis should not change the physics. We also remark that for boosts along the $z$-axis there is no azimuthal ($m$) mixing by symmetry of the problem. This enforces that $m_1 = m_2$ and the sum over $m_2$ drops out. We then obtain
\bealf{
\label{app:wigner_d}
^{d}_{s}\mathcal{K}_{\ell' \ell}^{m'm}(\boldsymbol{\beta})
= \sum_{m_1} D^{\ell'}_{m'm_1}(\phi,\theta,\psi) \bra{s \ell'm_1}\, e^{i\eta {\hat{Y}_z}}\,\ket{s \ell m_1}[D^{\ell}_{mm_1}(\phi,\theta,\psi)]^*.
}
This expression also follows from Eq.~(11) of \citep{Challinor2002} when setting $\vek{\varv}=\beta \hat{\vek{z}}$ and $R \vek{\varv} = \vb$. The relation between the Wigner D-functions and the spin-weighted spherical harmonics, $_sY_{lm}$, is 
\bealf{
\label{app:wigner_d_spherical}
D_{m m'}^\ell(\phi,\theta,\psi)&\equiv
(-1)^{m'}\sqrt{\frac{4\pi}{2\ell+1}}\,{_{-m'}}Y^*_{\ell m}(\theta, \phi)\,\expf{-im'\psi},
}
which allows us to rewrite Eq. (\ref{app:wigner_d}) as 
\bealf{
\label{app:kernel_elements_spherical}
^{d}_{s}\mathcal{K}_{\ell' \ell}^{m'm}(\boldsymbol{\beta})
=\sum_{m_1} \frac{4\pi\,{}_{-m_1}Y^*_{\ell'm'}(\vbh)\,\,
{}^{d}_{s}\mathcal{K}_{\ell' \ell}^{m_1}(\beta)\,\,
{}_{-m_1}Y_{\ell m}(\vbh)}{\sqrt{(2\ell + 1)(2\ell'+1)}},
}
which is our final result. Equation~\eqref{app:kernel_elements_spherical} allows one to calculate the aberration kernel elements for an arbitrary boost direction, $\vbh$, provided the elements for boosts collinear with the $z$-axis are known. Referring to current literature, \cite{Goldberg1967} and \cite{Hu1997} give the identity 
\bealf{
\label{app:wigner_d_spherical_Gold}
D^{\ell, \rm G}_{m' m}(\phi,\theta,\psi)
= \sqrt{\frac{4\pi}{2\ell + 1}}e^{i m'\psi}{_{-m'}}Y^{\rm G}_{\ell m}(\theta, \phi).
}
This has two differences with respect to the definitions we have employed. First, we have included the Condon-Shortley phase ${_{-s}}Y^{\rm G}_{\ell m}(\theta, \phi)=(-1)^{m}{_{-s}}Y_{\ell m}(\theta, \phi)$ which is consistent with the standard definition of {\tt Mathematica}. Second, in \cite{Goldberg1967} the Wigner D-matrix is given for the inverse rotation. This means 
\bealf{
\label{app:wigner_d_spherical_Gold_normal}
D^{\ell, \rm G}_{m'm}(\phi,\theta,\psi)
\equiv (-1)^{m-m'}\,\bra{\ell m'}\hat{U}^\dagger(R)\ket{\ell m}
= (-1)^{m-m'}\,\left[D^{\ell}_{m m'}(\phi,\theta,\psi)\right]^*,
}
which is consistent with Eq.~\eqref{app:wigner_d_spherical} implying our respective choices of convention also balance. This result provides an exact, fully general expression for the aberration kernel for arbitrary boost direction, $\vbh$. This form of the kernel is particularly useful for applications involving random electron velocities and although similar expressions have been given for the kernel for arbitrary $\vbh$ (see \cite{Challinor2002}, \cite{Dai2014}), to our knowledge this particular expression has not been reported.

\section{Boost operator}
\label{sec:BO}

\subsection{Introduction to the boost operator}
\label{sub:BO intro}
\noindent
In this section, the definition and properties of the boost operator are summarized. Whereas the aberration kernel encodes the transformation of the multipole coefficients under aberration for quantities which are independent of the photon frequency, $\nu$, the boost operator plays the same role for quantities which now depend on $\nu$. Again considering a boost along the $z$-axis, Eq.~\eqref{app:AK} generalizes to
\bealf{
\label{app:Boost operator}
{}_s X_{\ell m}'(\nu)
&= 
\sum_{\ell' m'} {}^{d}_{-s}\mathcal{\hat{B}}^{m}_{\ell \ell'}(\nu, \beta)\,{}_s X_{\ell'm}(\nu),
}
where ${}^{d}_{s}\mathcal{\hat{B}}^{m}_{\ell \ell'}(\nu, \beta)$ is the boost operator introduced by \citep{ChlubaBO25}. Explicitly, the boost operator can be expressed as
\bealf{
\label{eq:Boost op full}
{}^{d}_{s}\mathcal{\hat{B}}^{m}_{\ell \ell'}(\nu, \beta)
&=\int \text{d}\boldsymbol{\hat{n}}'\,{}_{-s}Y_{\ell m}^*(\boldsymbol{\hat{n}}')\left[\sum_{k=0}^{\infty}\frac{1}{k!}\left(\frac{\nu}{\nu'}\right)^{-d}\left[\ln{\left(\frac{\nu}{\nu'}\right)}\right]^k(-1)^k\oOnu^k\right]\,{}_{-s}Y_{\ell' m'}(\boldsymbol{\hat{n}}),
}
with the energy shift generator, $\oOnu=-\nu\partial_{\nu}$ and $\nu/\nu'=\gamma(1+\boldsymbol{\beta}\boldsymbol{\hat{n}}')$. Note the opposite sign convention relative to the aberration kernel, which favored the perspective of the observer. The operator $\oOnu$ is a dimensionless operator which generates infinitesimal rescalings of the energy. In our context, it tells us how the photon occupation number, $n(\nu)$, responds when the energy is rescaled due to the Doppler shift of the frequency. Note that taking the $k=0$ term only (i.e., no energy shift) would return us to the definition of the aberration kernel. 

As demonstrated in \cite{ChlubaBO25}, we can directly relate Eq.~\eqref{eq:Boost op full} to the aberration kernel. Absorbing the $(-1)^k$ into the logarithm and expressing what is left as an exponential, we have
\bealf{
\label{Boost op full 2}
{}^{d}_{s}\mathcal{\hat{B}}^{m}_{\ell \ell'}(\nu, \beta)
&=\int \text{d}\boldsymbol{\hat{n}}'\,{}_{-s}Y_{\ell m}^*(\boldsymbol{\hat{n}}')\left(\frac{\nu'}{\nu}\right)^{d}\text{exp}\left[\ln{\left(\frac{\nu'}{\nu}\right)}\,\oOnu\,\right]{}_{-s}Y_{\ell' m'}(\boldsymbol{\hat{n}})
\nonumber
\\
\nonumber
&=\int \text{d}\boldsymbol{\hat{n}}'\,{}_{-s}Y_{\ell m}^*(\boldsymbol{\hat{n}}')\left(\frac{\nu'}{\nu}\right)^{d+\oOnu}{}_{-s}Y_{\ell' m'}(\boldsymbol{\hat{n}})
\\
&=\int \text{d}\boldsymbol{\hat{n}}'\,\frac{{}_{-s}Y_{\ell m}^*(\boldsymbol{\hat{n}}'){}_{-s}Y_{\ell' m'}(\boldsymbol{\hat{n}})}{[\gamma\,(1+\beta\mu')]^{\,d+\oOnu}},
}
where we have used the commutation of $\oOnu$ with $\nu'/\nu$ (since $\nu'/\nu$ is independent of $\nu$ at fixed direction). We recognise the final expression in Eq.~\eqref{Boost op full 2} as the aberration kernel for a boost $-\beta$ along the $z$-axis, cf. Eq.~\eqref{eq:Aberration Kernel}. Thus we have proven the identity
\bealf{
\label{eq:boost_operator_z}
{}^{d}_{s}\hat{\mathcal{B}}^{m}_{\ell \ell'}(\nu,\beta)
\equiv {}^{d+\oOnu}_{s}\hat{\mathcal{K}}^{m}_{\ell \ell'}(-\beta),
}
as given in Eq. ($9$) of \cite{ChlubaBO25}. Note that here the operator $\oOnu$ now plays the role of a Doppler weight. Since rotations of the coordinate system commute with $\oOnu$, we then also have
\bealf{
\label{eq:boost_operator}
{}^{d}_{s}\hat{\mathcal{B}}^{mm'}_{\ell \ell'}(\nu,\boldsymbol{\beta})
\equiv {}^{d+\oOnu}_{s}\hat{\mathcal{K}}^{mm'}_{\ell \ell'}(-\boldsymbol{\beta}),
}
for general direction of the boost. This is the key result of this section and becomes crucial to the derivation that is to come.

\subsection{Symmetries of the boost operator}
\label{sub:BO symmetries}
\noindent
The symmetries of the aberration kernel summarized in Appendix~\ref{app:Kernel symmetries} can be extended to the boost operator by simple application of Eq.~(\ref{eq:boost_operator}). We state them here, briefly, to demonstrate our understanding of the boost operator's symmetry properties. First considering Eq. (\ref{app:Kernel relation 2}) we have
\bealf{
\label{eq:BO relation 2}
{}^{d}_{s}\hat{\mathcal{B}}_{\ell' \ell}^{m}\,(\nu, \beta)
\equiv{}^{d+\oOnu}_{s}\mathcal{K}_{\ell' \ell}^{m}\,(-\beta)
= 
(-1)^{\ell +\ell'}\,{}^{d+\oOnu}_{-s}\mathcal{K}_{\ell' \ell}^{m}\,(\beta)
\equiv
(-1)^{\ell +\ell'}\,{}^{\,d}_{-s}\hat{\mathcal{B}}_{\ell' \ell}^{m}\,(\nu, -\beta).
}
Thus for $s=0$, as will be the case we consider, there is a phase of $(-1)^{\ell+\ell'}$ needed to relate boost operators for boosts in anti-collinear directions. Next we extend Eq. (\ref{app:Aberration Kernel with s reversed}), finding that
\bealf{
\label{eq: BO with s reversed}
{}^{d}_{s}\hat{\mathcal{B}}_{\ell' \ell}^{m}\,(\beta)
\equiv
{}^{d+\oOnu}_{s}\mathcal{K}_{\ell' \ell}^{m}\,(-\beta)
= {}^{d+\oOnu}_{-s}\mathcal{K}_{\ell' \ell}^{-m}\,(-\beta)
\equiv {}^{d}_{-s}\hat{\mathcal{B}}_{\ell' \ell}^{-m}\,(\beta).
}
Again, for $s=0$ there is a full symmetry between boost operators for identical boosts but with inverted $m$. We can also develop Eq. (\ref{app:Kernel doppler relation}) and Eq. (\ref{app:Kernel doppler relation 2}), giving the final results
\bsub
\bealf{
\label{eq:BO doppler relation}
{}^{d}_{}\hat{\mathcal{B}}_{\ell' \ell}^{m}\,(\nu, \beta)
&= 
\gamma \,{}^{d+1}_{}\hat{\mathcal{B}}_{\ell' \ell}^{m}\,(\nu, \beta) +\gamma \beta \left[ C^{m}_{\ell'+1}\,{}^{d+1}_{}\hat{\mathcal{B}}_{\ell'+1 \ell}^{m}\,(\nu, \beta) + C^{m}_{\ell'}\,{}^{d+1}_{}\hat{\mathcal{B}}_{\ell'-1 \ell}^{m}\,(\nu, \beta)\right]
\\
&= 
\gamma \,{}^{d-1}_{}\hat{\mathcal{B}}_{\ell' \ell}^{m}\,(\nu, \beta) -\gamma \beta \left[ C^{m}_{\ell+1}\,{}^{d-1}_{}\hat{\mathcal{B}}_{\ell' \ell+1}^{m}\,(\nu, \beta) + C^{m}_{\ell}\,{}^{d-1}_{}\hat{\mathcal{B}}_{\ell' \ell-1}^{m}\,(\nu, \beta)\right].
}
\esub
The simplicity of Eq. (\ref{eq:boost_operator}) thus allows us to immediately evaluate the required boost operator elements given our knowledge of the aberration kernel elements. 

A key point to emphasize is that the aberration kernels, and thus the boost operators, are exact, all-order expressions in the boost velocity, $\beta$, for the transformation of photon field multipoles under Lorentz boosts. That is to say that unlike perturbative treatments the operator expressions remain completely valid. As an example, 
\bealf{
\label{eq: Kernel example}
{}^{d}_{}\hat{\mathcal{B}}_{0 0}^{0}\,(\nu, \beta)=
{}^{d+\oOnu}_{}\hat{\mathcal{K}}_{0 0}^{0}\,(-\beta)
=\frac{\left(\gamma-p\right)^{1-d-\oOnu}-\left(\gamma+p\right)^{1-d-\oOnu}}{2\left(d+\oOnu-1\right)p},
}
which gives the exact expression to generate expressions to all orders in $p$. Moreover, the number of symmetries and recursion relations attributed to the kernel allow any element to be obtained from a small number of basic elements which are calculated directly. Thus entire matrices of kernel elements can be generated efficiently, all without approximation, as is done in a \verb|Mathematica| notebook (linked in section \ref{Conclusion}). As a result of Eq. (\ref{eq:boost_operator}), this is also the case for the boost operator suggesting that a boost operator approach to CMB photon scattering problems can provide general, all-order expressions for lab frame collision terms in terms of specific boost operators.

\section{Rest frame collision term }
\label{sec: Rest frame col term}
\noindent
In this section, we derive the required expressions for the Compton collision term in the rest frame of the moving electron, including electron recoil terms up to first order.
We must focus on the Compton scattering process defined by  $e+\gamma_0 \rightarrow e' +\gamma$, where we define our initial, incoming photon, $\gamma_0$, and our final, outgoing photon, $\gamma$.
Throughout we use $\mu_{\rm{sc}}=\vgh_0\cdot\vgh$ to denote the direction cosine between the incoming and outgoing photon directions, $\vgh_0$ and $\vgh$, respectively. We also use the dimensionless energies $\omega_0=h \nu_0/\me c^2$ and $\omega=h \nu/\me c^2$, respectively. 
We will separately consider the effect of stimulated scattering events, $\propto n^2$, which are crucial for driving the photon occupation number $n(\omega, \vgh)$ towards full thermal equilibrium.

\subsection{Rest frame scattering cross section to first order in the electron recoil}
\label{sub:Rest frame recoil terms cross section}
\noindent
We shall begin by giving the Thomson terms and first-order recoil corrections of the scattering cross section. In the rest frame, the Compton scattering cross section\footnote{We scaled the cross section by $\sigT$ for convenience.} for the reaction is given by \citep[e.g.,][]{Jauch1976}
\begin{equation}
\label{eq:App_kerdef3}
\frac{\text{d}\sigma(\omega_0, \omega, \mu_{\rm{sc}})}{\text{d}\Omega} = \frac{3}{16\pi}\,\Bigg[\frac{\omega}{\omega_0}\Bigg]^2
\left[
1+\mu_{\rm{sc}}^2+
\left(\frac{\omega}{\omega_0} + \frac{\omega_0}{\omega}-2\right)\right],
\qquad
\frac{\omega}{\omega_0} = \frac{1}{1+\omega_0(1-\mu_{\rm{sc}})}.
\end{equation}
To obtain the classical Kompaneets equation, only first-order temperature corrections are needed, which means we can expand the cross section to first order in $\omega_0\ll 1$ to obtain
\bealf{
\label{app:first_order_omega}
\frac{\text{d}\sigma(\omega_0, \omega, \mu_{\rm{sc}})}{\text{d}\Omega} = \frac{3}{16\pi}(1+\mu_{\rm sc}^2)-\frac{3}{8\pi}\omega_0(1-\mu_{\rm sc})(1+\mu_{\rm sc}^2)+\mathcal{O}(\omega_0^2).
}
Here, we recognize the first group of terms as the Thomson scattering cross section, which neglects energy exchange (i.e., $\omega=\omega_0$).
The remaining terms account for the leading order electron recoil corrections, $\propto \omega_0$. Recoil refers to the fact that a small amount of energy is transferred to the electron as it acquires a momentum vector in some direction that is determined by the scattering angle. 

Omitting terms $\mathcal{O}(\omega_0^2)$, we can express the cross section as a Legendre polynomial expansion in terms of $\mu_{\rm sc}$, giving 
\bealf{
\label{app:LegendreP_first_order}
\frac{\text{d}\sigma(\omega_0, \omega, \mu_{\rm{sc}})}{\text{d}\Omega} \approx \frac{1}{4\pi}\Bigg[1+\frac{1}{2}P_2(\mu_{\rm sc})\Bigg] - \frac{2\omega_0}{4\pi}\sum_{\ell = 0}^3 \,(2\ell + 1)\,c_\ell P_\ell (\mu_{\rm sc}),
}
where $c_0 = 1$, $c_1 = -2/5$, $c_2 = 1/10$, $c_3 = -3/70$. At $\mathcal{O}(\omega_0)$, the expansion truncates at $\ell=3$ as higher multipoles are suppressed by higher orders of $\omega_0$. Using the addition theorem for spherical harmonics, $P_{\ell}\,(\mu_{\rm sc})=\frac{4\pi}{2\ell+1} \sum^{\ell}_{m=-\ell} Y_{\ell m}(\vgh_0)Y_{\ell m}^*(\vgh)$, we rewrite the cross section as 
\bealf{
\label{app:Spherical_first_order}
\frac{\text{d}\sigma(\omega_0, \omega, \mu_{\rm{sc}})}{\text{d}\Omega} \approx Y_{00}(\vgh_0)\,Y^*_{00}(\vgh)+\frac{1}{10}\sum_{m=-2}^2
Y_{2m}(\vgh_0)\,Y^*_{2m}(\vgh) - 2\omega_0 \sum_{\ell=0}^3 \, c_\ell \sum_{m=-\ell}^\ell Y_{\ell m}(\vgh_0)Y^*_{\ell m}(\vgh),
}
obtaining the coordinate independent form. This is the Compton scattering cross section to first order in $\omega_0$ in the electron rest frame. 

\subsection{Rest frame collision term without stimulated scattering}
\label{sub:Rest frame recoil terms}
\noindent
We next define the Compton collision term in the electron frame \citep{Buchler1976, Sazonov2000}
\begin{align}
\label{app:col_term_int}
\frac{\text{d}n(\omega_0, \vgh_0)}{\text{d}\tau}  
&= \int \text{d}\vgh\,
\Bigg\{\frac{\text{d}\sigma(\omega_0, \tilde{\omega},\mu_{\rm{sc}})}{\text{d}\Omega}\,n(\tilde{\omega}, \vgh)[1+n(\omega_0, \vgh_0)]
\nonumber \\
&\qquad \qquad \qquad \qquad - \frac{\text{d}\sigma(\omega_0, \omega, \mu_{\rm{sc}})}{\text{d}\Omega}\,n(\omega_0, \vgh_0)[1+n(\omega, \vgh)]\Bigg\},
\end{align}
where $\tilde{\omega}=\omega_0/[1-\omega_0(1-\mu_{\rm sc})]$, as required to ensure that the scattered photon has energy $\omega_0$ when running the process backwards for the first term in the integral.\footnote{We confirmed by brute force computation that this indeed yields the correct final result up to first-order temperature corrections and including stimulated terms \citep[see][]{Sazonov2000}.} The collision term describes how the photon occupation number at a given phase-space point $(\omega_0, \vgh_0)$ changes with optical depth, $\tau$. We are effectively counting the number of photons which scatter into the $(\omega_0, \vgh_0)$ state from some state $(\tilde{\omega}, \vgh)$ [the gain (positive) term], and subtract those photons which scatter from $(\omega_0, \vgh_0)$ into some final state $(\omega, \vgh)$ [the loss (negative) term]. We must integrate over all possible photon directions, $\vgh$, of the scattered photon. The $[1+n(\omega_0, \vgh_0)]$ and $[1+n(\omega, \vgh)]$ Bose enhancement factors account for the fact that a photon, as a boson, is more likely to scatter into a state with a high occupation number. 

To continue, for now, we neglect stimulated scattering terms $\propto n^2$, but will return to those in section \ref{sub:Rest stim terms}. 
We also note that
\begin{align}
\label{app:ds_in}
\frac{\text{d}\sigma(\omega_0, \tilde{\omega},\mu_{\rm{sc}})}{\text{d}\Omega}&\approx Y_{00}(\vgh_0)\,Y^*_{00}(\vgh)+\frac{1}{10}\sum_{m=-2}^2
Y_{2m}(\vgh_0)\,Y^*_{2m}(\vgh) + 2\omega_0 \sum_{\ell=0}^3 \, c_\ell \sum_{m=-\ell}^\ell Y_{\ell m}(\vgh_0)Y^*_{\ell m}(\vgh)
\nonumber
\\
&\approx \frac{\text{d}\sigma(-\omega_0, \omega,\mu_{\rm{sc}})}{\text{d}\Omega},
\end{align}
which simplifies matters significantly. 
In our gain term, since $\tilde{\omega}$ is a function of $\omega_0$, we must expand $n(\tilde{\omega},\vgh)$ about $\omega_0$, up to first order in $\omega_0$. We find 
\bealf{
\label{app:photon_occ_expansion}
n(\tilde{\omega},\vgh) \approx n(\omega_0,\vgh) - \omega_0 (1- \mu_{\rm sc}) \,\hat{\mathcal{O}}_{\omega_0}n(\omega_0, \vgh),
}
where $\hat{\mathcal{O}}_{\omega_0}\equiv-\omega_0 \partial_{\omega_0}$.\footnote{Note that since the energy shift generator is dimensionless, $\hat{\mathcal{O}}_{\omega}\equiv\oOnu\equiv \hat{\mathcal{O}}_{x}$ etc..}
Substituting this back into Eq. (\ref{app:col_term_int}) [after dropping terms $\propto n^2$] and performing the angular integrals we finally obtain
\bealf{
\label{app:collision_term_first_order}
\frac{\text{d}n(\omega_0, \vgh_0)}{\text{d}\tau}  
\approx n_0 + \frac{1}{10}n_2-n
+2\omega_0\Bigg[\left(1-\frac{1}{2}\hat{\mathcal{O}}_{\omega_0}\right)\left(n_0 -\frac{2}{5}n_1+ \frac{1}{10}n_2 - \frac{3}{70}n_3\right) + n\Bigg].
}
Here, we introduced $n_\ell=\sum_{m=-\ell}^\ell Y_{\ell m}(\vgh_0)\,n_{\ell m}(\omega_0)$ and have suppressed the $(\omega_0, \vgh_0)$ arguments for convenience. The first group of terms is the Thomson result which has no recoil corrections. The terms $2\omega_0[n_0 -\frac{2}{5}n_1+ \frac{1}{10}n_2 - \frac{3}{70}n_3 + n]$ derive from recoil corrections to the total cross section, while those $\propto \hat{\mathcal{O}}_{\omega_0}$ are related to changes in the photon energy of $\tilde{\omega}\rightarrow \omega_0$. 

\subsection{Rest frame collision term from stimulated scattering}
\label{sub:Rest stim terms}
\noindent
We now return to the simulated scattering terms $\propto n^2$ in the collision term, previously neglected. Once again, we consider them to first order in $\omega_0$. The neglected terms from Eq. (\ref{app:col_term_int}) are the following,
\begin{align}
\label{app:stim_terms}
\frac{\text{d}n(\omega_0, \vgh_0)}{\text{d}\tau}\Bigg|^{n^2}
&= \int \text{d}\vgh\,
\Bigg\{\frac{\text{d}\sigma(\omega_0, \tilde{\omega},\mu_{\rm{sc}})}{\text{d}\Omega}\,n(\tilde{\omega}, \vgh)\,n(\omega_0, \vgh_0)- \frac{\text{d}\sigma(\omega_0, \omega, \mu_{\rm{sc}})}{\text{d}\Omega}\,n(\omega_0, \vgh_0)\,n(\omega, \vgh)\Bigg\}. 
\end{align}
First considering only the gain term, the multiplicative factor $n(\omega_0, \vgh_0)$ carries through the integral, essentially giving each previously calculated gain term multiplied by $n(\omega_0, \vgh_0)$. Explicitly we have,
\begin{align}
\label{app:first_stim_term}
\frac{\text{d}n(\omega_0, \vgh_0)}{\text{d}\tau}\Bigg|^{n^2}_{\text{gain}} 
\approx n\Bigg[ n_0 + \frac{1}{10}n_2
+2\omega_0\left(1-\frac{1}{2}\hat{\mathcal{O}}_{\omega_0}\right)\left(n_0 -\frac{2}{5}n_1+ \frac{1}{10}n_2 - \frac{3}{70}n_3\right) \Bigg],
\end{align}
again, suppressing the arguments. However, the neglected loss term is not as straightforward. As before, the factor $n(\omega, \vgh)$ must be expanded in powers of $\omega_0$ similar to Eq.~\eqref{app:photon_occ_expansion}, giving
\bealf{
\label{app:photon_occ_expansion_omega}
n(\omega,\vgh) \approx n(\omega_0,\vgh) + \omega_0 (1- \mu_{\rm sc}) \,\hat{\mathcal{O}}_{\omega_0}n(\omega_0, \vgh).
}
This means that we encounter the same integrals as for the gain term, but with a sign-flip in the recoil contributions. We thus can immediately write
\begin{align}
\label{app:second_stim_term}
\frac{\text{d}n(\omega_0, \vgh_0)}{\text{d}\tau}\Bigg|^{n^2}_{\text{loss}}
\approx -n\Bigg[n_0+\frac{1}{10}n_2-2\omega_0\left(1-\frac{1}{2}\hat{\mathcal{O}}_{\omega_0}\right)\left(n_0-\frac{2}{5}n_1+\frac{1}{10}n_2-\frac{3}{70}n_3\right)\Bigg].
\end{align}
Combining both terms compactly, we find
\begin{align}
\label{app:full_stim_term}
\frac{\text{d}n(\omega_0, \vgh_0)}{\text{d}\tau}\Bigg|^{n^2}
\approx 4\omega_0\,n\,\left(1-\frac{1}{2}\hat{\mathcal{O}}_{\omega_0}\right)\Bigg(n_0 - \frac{2}{5}n_1+\frac{1}{10}n_2-\frac{3}{70}n_3\Bigg),
\end{align}
again suppressing the $(\omega_0, \vgh_0)$ arguments. Note that the corrections from stimulated terms have no zeroth order $\omega_0$ term as expected since in the Thomson limit the stimulated terms vanish identically due to the lack of energy exchange. Note also that we now have terms non-linear in $n$. 

\subsection{Final expression for the rest frame collision term}
\label{sub:Rest stim terms all}
\noindent
By combining all terms given above, we find the rest frame Compton collision term at first order in the electron recoil as 
\bealf{
\label{app:collision_term_first_order_all}
\frac{\text{d}n(\omega_0, \vgh_0)}{\text{d}\tau}  
\approx n_0 + \frac{1}{10}n_2-n
+2\omega_0\Bigg[(1+2n)\left(1-\frac{1}{2}\hat{\mathcal{O}}_{\omega_0}\right)\left(n_0 -\frac{2}{5}n_1+ \frac{1}{10}n_2 - \frac{3}{70}n_3\right) + n\Bigg].
}
The non-linear terms in the photon occupation number are crucial for obtaining the final Kompaneets equation and driving the photon distribution towards a Bose-Einstein spectrum. However, these terms also lead to a complicated scattering structure in the presence of photon anisotropies as we shall see below. As we demonstrate, at lowest order in the electron temperature, this scattering structure can be captured using the Gaunt integrals without any new effects coming from the boosts themselves. 

\section{Lab frame collision term }
\label{sec:Lab frame col term}

\subsection{Lab frame expression without stimulated scattering effects}
\label{sub:Thomson and leading order recoil terms}
\noindent
We now give the lab frame collision terms including all orders in the electron momenta up to first order in the electron recoil. Neglecting stimulated scattering corrections, we have 
\bealf{
\label{app:collision_term}
\frac{\text{d}n'(\omega', \vgh')}{\text{d}\tau'}
\approx 
n_0' + \frac{1}{10}n_2'-n'+2\omega'\left[
\left(1-\frac{\hat{\mathcal{O}}_{\omega'}}{2}\right)\left(n_0'-\frac{2}{5}n_1'+\frac{1}{10}n_2'-\frac{3}{70}n_3'\right)+n'\right],
}
in the electron rest frame. With the boost operator we can obtain the spherical harmonic coefficients in this frame using 
\bealf{
\label{app:coefficient_transform}
n_{\ell m}'(\omega')=\sum_{\ell'}{}^{0}\hat{\mathcal{B}}^m_{\ell \ell'}(\omega',\beta)\,n_{\ell'm}(\omega'),
}
where we have assumed that the boost is along the $z$-axis. Inserting this into the expression for the collision term we then can evaluate the scattering operation to obtain $\text{d}n'(\omega', \vgh')/\text{d}\tau'$. We can next transform back into the lab frame using a boost in the opposite direction. For this we must also consider the Lorentz transformation of the optical depth which is given by\footnote{$\text{d}\tau'=n_e'\,\sigma_T'\,\text{d}t'=(n_e/\gamma)\,\sigma_T\,\text{d}t\,\gamma(1-\beta\mu_{}) = \text{d}\tau\,(1-\beta\mu_{})$, for a boost along $-\boldsymbol{\hat{z}}$.} \citep[][]{Chluba2012SZpack, ChlubaBO25}
\bealf{
\label{app:optical_depth_transform}
\text{d}\tau'=(1-\beta\mu)\,\text{d}\tau=\Bigg(\frac{\omega'}{\omega}\Bigg)\,\frac{\text{d}\tau}{\gamma},
}
where $\mu\equiv\cos\theta$ is the direction cosine, for a transformation in the $-\boldsymbol{\hat{z}}$ direction. This implies that the harmonic coefficients of $\text{d}n'(\omega', \vgh')/\text{d}\tau'$ transform with a Doppler-weight that is reduced by one and picks up an overall factor of $1/\gamma$ \citep[see also][]{ChlubaBO25, Chluba2026SZ}. 
Due to invariance of the photon phase-space distribution, we have $n(\omega, \vgh)= n'(\omega'[\omega, \vgh], \vgh'[\vgh])$. This allows us to get the transformed collision term in the lab frame for a boost along $-\boldsymbol{\hat{z}}$ using
\bealf{
\label{app:col_term_depth_transform}
\frac{\text{d}n(\omega, \vgh)}{\text{d}\tau}=\frac{1}{\gamma}
\Bigg(\frac{\omega'}{\omega}\Bigg)\,\frac{\text{d}n'(\omega'[\omega, \vgh], \vgh'[\vgh])}{\text{d}\tau'}.
}
To give the final expression, we have to consider how each of the terms in Eq.~\eqref{app:collision_term} transforms. The Thomson terms all have a Doppler-weight $d=0$ and thus transform as ${}^{-1}\hat{\mathcal{B}}^m_{\ell \ell'}(\omega,-\beta)/\gamma$ once accounting for the factor from the optical depth. The operator $\hat{\mathcal{O}}_{\omega'}$ is Lorentz-invariant, meaning $\hat{\mathcal{O}}_{\omega'}\rightarrow \hat{\mathcal{O}}_{\omega_0}$ without further ado. The terms $\propto \omega' n'$ have a Doppler-weight $d=-1$, as they can be written as $\omega' n' = \omega \, (\omega'/\omega) \,n'$. The terms $(\omega'/\omega) \,n'$ then transform using ${}^{-2}\hat{\mathcal{B}}^m_{\ell \ell'}(\omega,-\beta)/\gamma$. 

With the rules discussed above, we can in principle already write the desired final expression. However, we notice that we encounter the terms ${}^{-1}\hat{\mathcal{B}}^m_{\ell \ell'}(\omega,-\beta)/\gamma$ and ${}^{-2}\hat{\mathcal{B}}^m_{\ell \ell'}(\omega,-\beta)/\gamma$ always in combination with ${}^{0}\hat{\mathcal{B}}^m_{\ell' \ell}(\omega,\beta)$ from the transformation into the moving frame. As a result of this repeating structure, it is therefore useful to define the Doppler operator,
\bealf{
\label{app:D}
^d\hat{\mathcal{D}}^m_{\ell\ell'\ell''}(\omega,\beta)\equiv\frac{{}^{d}\hat{\mathcal{B}}^m_{\ell \ell'}(\omega,-\beta)\,{}^{0}\hat{\mathcal{B}}^m_{\ell' \ell''}(\omega,\beta)}{\gamma},
}
to describe the transformation from the initial $\ell''$ to the final $\ell$ multipole via the intermediate state $\ell'$ which depends on the angular dependence of the scattering process. With this, we can then directly write the final expression for the transformed collision term as
\bealf{
\label{app:collected_sums}
\frac{\text{d}n(\omega, \vgh)}{\text{d}\tau}
&= \sum_{\ell m \ell''}Y_{\ell m}(\vgh)\Bigg\{{}^{-1}\hat{\mathcal{D}}^m_{\ell0\ell''}+\frac{1}{10}{}^{-1}\hat{\mathcal{D}}^m_{\ell2\ell''}-\sum_{\ell'}{}^{-1}\hat{\mathcal{D}}^m_{\ell\ell'\ell''} \nonumber
\\ \nonumber 
&\qquad +2\omega \,\Bigg[\left(1-\frac{\hat{\mathcal{O}}_{\omega}}{2}\right)\Bigg({}^{-2}\hat{\mathcal{D}}^m_{\ell0\ell''}-\frac{2}{5}{}^{-2}\hat{\mathcal{D}}^m_{\ell1\ell''}+\frac{1}{10}{}^{-2}\hat{\mathcal{D}}^m_{\ell2\ell''}-\frac{3}{70}{}^{-2}\hat{\mathcal{D}}^m_{\ell3\ell''}\Bigg)\\  
&\qquad\qquad+\sum_{\ell'}{}^{-2}\hat{\mathcal{D}}^m_{\ell\ell'\ell''}
\Bigg]\Bigg\}\,n_{\ell''m}(\omega),
}
which is exact to {\it all} orders in $p$ and up to first order in electron recoil. Importantly, one can directly arrive at this result from Eq.~\eqref{app:collision_term} by simply asking which Doppler-weight the required quantities have once the optical depth transformation is accounted for. The Doppler operator, also used in \citep{Chluba2026SZ} and \citep{Rosenberg2025prSZ} for $d=-1$, handles all the related transformations.

The expression in Eq.~\eqref{app:collected_sums} can be further simplified by using the identities
\bsub
\bealf{
\label{app:D_sum_over_l_prime}
\sum_{\ell'}{}^{-1}\hat{\mathcal{D}}^m_{\ell\ell'\ell''}&=\delta_{\ell \ell''}-\beta\Bigg[C^{m}_{\ell+1}\delta_{\ell+1 \ell''}+C^{m}_{\ell}\delta_{\ell-1 \ell''}\Bigg]
,
\\
\label{app:raise_d_of_D}
\sum_{\ell'}{}^{-2}\hat{\mathcal{D}}^m_{\ell\ell'\ell''}
&=\sum_{\ell'}\gamma
\,\Bigg\{{}^{-1}\hat{\mathcal{D}}^m_{\ell\ell'\ell''}-\beta\Bigg[C^{m}_{\ell+1}{}^{-1}\hat{\mathcal{D}}^m_{\ell+1\ell'\ell''}+C^{m}_{\ell}{}^{-1}\hat{\mathcal{D}}^m_{\ell-1\ell'\ell''}\Bigg]\Bigg\}
\nonumber\\
&=\gamma\,\delta_{\ell \ell''}-2\beta\gamma\Bigg[C^{m}_{\ell+1}\delta_{\ell+1 \ell''}+C^{m}_{\ell}\delta_{\ell-1 \ell''}\Bigg]
\nonumber
\\
&\qquad
+\beta^{2}\gamma\Bigg[C^{m}_{\ell+2}C^{m}_{\ell+1}\delta_{\ell+2\ell''}
+\Big\{(C^{m}_{\ell+1})^{2}+(C^{m}_{\ell})^{2}\Big\}\delta_{\ell\ell''}+C^{m}_{\ell}C^{m}_{\ell-1}\delta_{\ell-2\ell''}\Bigg],
}
\esub
where $C_\ell^m=\sqrt{(\ell^2-m^2)/(4\ell^2-1)}$. Here, we have used Eq.~(\ref{eq:BO doppler relation}) repeatedly in the second equation and the applied the important identity $\sum_{\ell'} {}^{d}\hat{\mathcal{B}}^m_{\ell \ell'}(\omega,-\beta)\,{}^{d}\hat{\mathcal{B}}^m_{\ell' \ell''}(\omega,\beta) = \delta_{\ell \ell''}$, which signifies that a boost followed by an identical boost in the opposite direction yields the identity \citep{ChlubaBO25}.

Applying these relations, the final result for the fully-simplified Compton collision term to first order in $\omega$ then reads
\bealf{
\label{app:fully_simplified}
\frac{\text{d}n(\nu, \vgh)}{\text{d}\tau}
&= \sum_{\ell m \ell''}Y_{\ell m}(\vgh)\left\{{}^{-1}\hat{\mathcal{D}}^m_{\ell0\ell''}+\frac{1}{10}{}^{-1}\hat{\mathcal{D}}^m_{\ell2\ell''}-\delta_{\ell \ell''}+\beta\Bigg[C^{m}_{\ell+1}\delta_{\ell+1 \ell''}+C^{m}_{\ell}\delta_{\ell-1 \ell''}\Bigg]\right\}
\,n_{\ell''m}(\nu)
\nonumber\\
&\qquad
+2\omega\sum_{\ell m \ell''}Y_{\ell m}(\vgh)
\Bigg\{
\left(1-\frac{\oOnu}{2}\right)\left(
{}^{-2}\hat{\mathcal{D}}^m_{\ell0\ell''}-\frac{2}{5}{}^{-2}\hat{\mathcal{D}}^m_{\ell1\ell''}+\frac{1}{10}{}^{-2}\hat{\mathcal{D}}^m_{\ell2\ell''}-\frac{3}{70}{}^{-2}\hat{\mathcal{D}}^m_{\ell3\ell''}\right)
\nonumber
\\
&\qquad\qquad
+\gamma\delta_{\ell\ell''}-2\beta\gamma\Bigg[C^{m}_{\ell+1}\delta_{\ell+1\ell''}+C^{m}_{\ell}\delta_{\ell-1\ell''}\Bigg]
+\beta^{2}\gamma\Bigg[C^{m}_{\ell+2}C^{m}_{\ell+1}\delta_{\ell+2\ell''}
\nonumber
\\
&\qquad\qquad\qquad
+\Big\{(C^{m}_{\ell+1})^{2}+(C^{m}_{\ell})^{2}\Big\}\,\delta_{\ell\ell''}
+C^{m}_{\ell}C^{m}_{\ell-1}\delta_{\ell-2\ell''}\Bigg]
\Bigg\}\, n_{\ell''m}(\nu),
}
where we have returned to using $\nu$ as the main variable. The first term in braces is the result for Thomson scattering without recoil included \citep[cf.,][]{ChlubaBO25}, where the terms multiplied by $\beta$ account for Doppler and aberration effects. All terms multiplying the factor of $2\omega$ account for first-order recoil effects with those terms proportional to both $\beta$ (or $\beta^2$) and $\omega$ characterized as mixed recoil and velocity terms. These final terms include mixing between Doppler effects, aberration and of course recoil. The key point is that Eq. (\ref{app:fully_simplified}), the main result of this section, retains the full Lorentz structure of the Doppler operators (to all orders in $p$) while expanding only in the recoil parameter $\omega$. 

\subsection{General direction collision term without stimulated scattering}
\label{sub:collision_term_gen_dir}
\noindent
The result given by Eq.~\eqref{app:fully_simplified} is valid for the specific case where we have $\vbh = \boldsymbol{\hat{z}}$. It is useful to generalise it to arbitrary directions which we can do using a similar procedure to that used by \cite{Rosenberg2025prSZ} and also as given in section~\ref{Arbitrary boost direction}. We start by writing Eq.~\eqref{app:fully_simplified} in a more compact form 
\bealf{
\label{app:z_dir_col_term}
\frac{\text{d}n(\nu, \vgh)}{\text{d}\tau}\Bigg|_{\vbh = \boldsymbol{\hat{z}}} 
&= \sum_{\ell m \ell''}Y_{\ell m}(\vgh)\left(\hat{\mathcal{S}}^m_{\ell\ell''}+2\omega\, \hat{\mathcal{R}}^m_{\ell\ell''}\right)n_{\ell''m}(\nu),
}
where we have condensed the Thomson and recoil terms into two operators $\hat{\mathcal{S}}^m_{\ell\ell''}$ and $ \hat{\mathcal{R}}^m_{\ell\ell''}$, respectively. Using the same concepts as in section \ref{Arbitrary boost direction}, we look to rotate into the $\vbh = \boldsymbol{\hat{z}}$ frame, perform the scattering and then rotate back into the general frame. Making use of our developed knowledge of such transformations, we recognize Wigner-D functions appearing as in Eq.~(\ref{app:wigner_d}). The collision term for general direction $\vbh$ is therefore
\bealf{
\label{app:gen_dir_col_term}
\frac{\text{d}n(\nu, \vgh)}{\text{d}\tau}\Bigg|_{\vbh} 
&= \sum_{\ell m
\ell''}Y_{\ell m}(\vgh)\sum_{m'' \tilde{m}}D^{\ell}_{m \tilde{m}}(\vbh)\left(\hat{\mathcal{S}}^{\tilde{m}}_{\ell\ell''}+2\omega \,\hat{\mathcal{R}}^{\tilde{m}}_{\ell\ell''}\right)[D^{\ell''}_{m''\tilde{m}}(\vbh)]^*n_{\ell''m''}(\nu).
}
Note the appearance of the two sums over $m''$ and $\tilde{m}$ which reintroduce the azimuthal mixing that previously vanished due to the condition that $\vbh$ and $\hat{\boldsymbol{z}}$ were collinear.

For our scattering problems, we assume that the electron distribution is isotropic and thermal. Therefore, we next perform an angle average of the general expression over all directions of $\vbh$.
The Wigner-D functions $D^{\ell}_{m \tilde{m}}(\vbh)$, $[D^{\ell''}_{m''\tilde{m}}(\vbh)]^*$ can be expressed in terms of spin-weighted spherical harmonics as in Eq.~\eqref{app:wigner_d_spherical}. Averaging over all boost directions, $\vbh$, we then find
\bealf{
\label{app:dir_avg_wigner_d}
\int \frac{\text{d}\vbh}{4\pi}\,D^{\ell}_{m \tilde{m}}(\vbh)\,[D^{\ell''}_{m''\tilde{m}}(\vbh)]^*=\int \frac{\text{d}\vbh}{4\pi}\frac{4\pi\,{_{-\tilde{m}}}Y^*_{\ell m}(\vbh)\,{_{-\tilde{m}}}Y_{\ell'' m''}(\vbh)}{\sqrt{(2\ell+1)(2\ell''+1)}}
= \frac{\delta_{\ell \ell''}\delta_{m m''}}{2\ell + 1}.
}
Applying this to our direction-averaged general collision term, we obtain
\bealf{
\label{app:avgd_gen_dir_col_term}
\left<\frac{\text{d}n(\nu, \vgh)}{\text{d}\tau}\Bigg|_{\vbh}\,\right>
&= \sum_{\ell m}\frac{Y_{\ell m}(\vgh)}{2\ell+1}\sum_{\tilde{m}}\left(\hat{\mathcal{S}}^{\tilde{m}}_{\ell\ell}+2\omega\, \hat{\mathcal{R}}^{\tilde{m}}_{\ell\ell}\right)n_{\ell m}(\nu).
}
This result is much needed for studying the scattering of anisotropies in the photon field. For the study of isotropic scattering, there is a simplification we can make which will cut out the need for this more general result (see section \ref{Kompaneets Equation}). 

Note that averaging over all $\vbh$ eliminates mixing between multipoles, enforcing $\ell=\ell''$ and $m=m''$. Since we have an intermediate sum over $\tilde{m}$, we can further simplify the overall operator expression. Substituting back in the expressions for $\hat{\mathcal{S}}^m_{\ell\ell}$ and $\hat{\mathcal{R}}^m_{\ell\ell}$, we have
\bealf{
\label{app:full_dir_avgd_col_term}
\left<\frac{\text{d}n(\nu, \vgh)}{\text{d}\tau}\Bigg|_{\vbh}\,\right>
&= \sum_{\ell m}Y_{\ell m}(\vgh)\left\{{}^{-1}\hat{\mathcal{D}}_{\ell0\ell}+\frac{1}{10}{}^{-1}\hat{\mathcal{D}}_{\ell2\ell}-1\right\}
\,n_{\ell m}(\nu)
\nonumber\\
&\qquad 
+2\omega\sum_{\ell m}Y_{\ell m}(\vgh)
\Bigg\{
\left(1-\frac{\oOnu}{2}\right)\left(
{}^{-2}\hat{\mathcal{D}}_{\ell0\ell}-\frac{2}{5}{}^{-2}\hat{\mathcal{D}}_{\ell1\ell}+\frac{1}{10}{}^{-2}\hat{\mathcal{D}}_{\ell2\ell}-\frac{3}{70}{}^{-2}\hat{\mathcal{D}}_{\ell3\ell}\right)
\nonumber
\\
&\qquad\qquad
+\gamma\left(1+\frac{\beta^{2}}{3}\right)
\Bigg\}\, n_{\ell m}(\nu),
}
where we have introduced ${}^{d}\hat{\mathcal{D}}_{\ell\ell'\ell''} = \sum_{\tilde{m}}\,{}^{d}\hat{\mathcal{D}}^{\tilde{m}}_{\ell\ell'\ell''}/(2\ell+1)$ as the $m$-averaged Doppler operator which must satisfy $|\tilde{m}| \leq \textrm{min}[\ell, \ell', \ell'']$. We have also used $\sum_m [(C^{m}_{\ell+1})^{2}+(C^{m}_{\ell})^{2}]=(2\ell+1)/3$ in the last term.

\subsection{Lab frame stimulated scattering effects}
\label{sub:Lab frame stimulated terms}
\noindent
We now turn our attention to the rest frame stimulated terms derived in section \ref{sub:Rest stim terms} and look to transform them to the lab frame. From Eq. (\ref{app:full_stim_term}) we have
\begin{align}
\label{app:rest_stim_term}
\frac{\text{d}n'(\nu', \vgh')}{\text{d}\tau'}\Bigg|^{n^2}
\approx 4\omega'\,n'\,\left(1-\frac{\hat{\mathcal{O}}_{\omega'}}{2}\right)\sum_{
\ell=0}^{3}c_\ell \,n'_\ell,
\end{align}
where, as before, $c_0 = 1$, $c_1 = -2/5$, $c_2 = 1/10$ and $c_3=-3/70$. Our first step is to express each term as its spherical harmonic expansion and use orthogonality properties as follows:
\begin{align}
\label{app:rest_stim_term_1}
\frac{\text{d}n'_{\ell m}(\nu')}{\text{d}\tau'}\Bigg|^{n^2}
&\approx 4\omega' \int\text{d}\vgh'Y_{\ell m}^*(\vgh')\left(\sum_{\ell_1m_1}n_{\ell_1m_1}'(\nu')Y_{\ell_1m_1}(\vgh')\right)\left(1-\frac{\hat{\mathcal{O}}_{\omega'}}{2}\right)\left(\sum_{\ell_2 m_2} c_{\ell_2} n_{\ell_2 m_2}'(\nu')Y_{\ell_2 m_2}(\vgh')\right)
\nonumber\\
&= 4\omega' \sum_{\ell_1 m_1}\sum_{\ell_2 m_2} c_{\ell_2}\,(-1)^m\mathcal{G}^{\ell,\ell_1,\ell_2}_{-m,m_1,m_2}n_{\ell_1 m_1}'(\nu')\left(1-\frac{1}{2}\hat{\mathcal{O}}_{\omega'}\right)n_{\ell_2 m_2}'(\nu').
\end{align}
In the last step, we expressed the integral over three spherical harmonic functions using the Gaunt integral (see Appendix~\ref{app:Gaunt}).\footnote{Since one of the harmonics is conjugated, we can use the symmetry $Y_{\ell m}^*=(-1)^m\,Y_{\ell, -m}$ which gives the required Gaunt coefficient as $(-1)^m \mathcal{G}^{\ell,\ell_1,\ell_2}_{-m,m_1,m_2}$.} Reverting back, we find 
\begin{align}
\label{app:rest_stim_term_ntilde}
\frac{\text{d}n'(\nu', \vgh')}{\text{d}\tau'}\Bigg|^{n^2}
&\approx 4\omega'\sum_{\ell m}Y_{\ell m}(\vgh')\sum_{\ell_1 m_1}\sum_{\ell_2 m_2} \,(-1)^mc_{\ell_2}\,\mathcal{G}^{\ell,\ell_1,\ell_2}_{-m,m_1,m_2}n_{\ell_1 m_1}'(\nu')\left(1-\frac{1}{2}\hat{\mathcal{O}}_{\omega'}\right)n_{\ell_2 m_2}'(\nu')
\end{align}
for the stimulated recoil contribution to the collision term in the electron rest frame. 

To obtain the expression in the lab frame, we have to transform the variable
\begin{align}
\label{app:ntilde}
\mathcal{N}_{\ell m}(\nu)
&
=
\sum_{\ell_1 m_1}\sum_{\ell_2 m_2} (-1)^mc_{\ell_2}\,\mathcal{G}^{\ell,\ell_1,\ell_2}_{-m,m_1,m_2}n_{\ell_1 m_1}(\nu)\left(1-\frac{1}{2}\hat{\mathcal{O}}_{\omega}\right)n_{\ell_2 m_2}(\nu),
\end{align}
which is an effective quantity arising from the non-linear combination of two photon occupation numbers with the energy shift operator. This quantity can be considered as an effective photon occupation number, which transforms with Doppler weight $d=0$ since $\hat{\mathcal{O}}_{\omega}$ is already Lorentz invariant.\footnote{The naive replacement $n'\rightarrow \mathcal{B}\, n$ is not valid in this case, as it does not capture the correct spherical harmonic couplings.} 
Now transforming all terms as we usually would, we have 
\begin{align}
\label{app:rest_stim_term_ntilde_transformed}
\frac{\text{d}n(\nu, \vgh)}{\text{d}\tau}\Bigg|^{n^2}
&\approx 4\omega\sum_{\ell \tilde{m}}Y_{\ell \tilde{m}}(\vgh)\sum_{\ell' \tilde{\ell}}{}^{-2}\hat{\mathcal{D}}_{\ell \ell' \tilde{\ell}}^{\tilde{m}}(\nu, \beta)\,\mathcal{N}_{\tilde{\ell} \tilde{m}}(\nu).
\end{align}
Inserting Eq.~(\ref{app:raise_d_of_D}) for the intermediate $\ell'$ summation, we obtain
\begin{align}
\label{app:Insert_Dopp_identity}
\frac{\text{d}n(\nu, \vgh)}{\text{d}\tau}\Bigg|^{n^2}
&\approx 4\gamma \omega \sum_{\ell \tilde{m}}Y_{\ell \tilde{m}}(\vgh)\sum_{\tilde{\ell}}\Bigg[\delta_{\ell \tilde{\ell}}-2\beta\,\Bigg(C^{\tilde{m}}_{\ell+1}\delta_{\ell+1, \tilde{\ell}}+C^{\tilde{m}}_{\ell}\delta_{\ell-1, \tilde{\ell}}\Bigg)
\nonumber
\\
&\qquad
+\beta^{2}\,\Bigg(C^{\tilde{m}}_{\ell+2}C^{\tilde{m}}_{\ell+1}\delta_{\ell+2,\tilde{\ell}}
+\Big\{(C^{\tilde{m}}_{\ell+1})^{2}+(C^{\tilde{m}}_{\ell})^{2}\Big\}\delta_{\ell\tilde{\ell}}+C^{\tilde{m}}_{\ell}C^{\tilde{m}}_{\ell-1}\delta_{\ell-2,\tilde{\ell}}\Bigg)\Bigg]\mathcal{N}_{\tilde{\ell} \tilde{m}}(\nu).
\end{align}
Unlike in previous treatments, the stimulated recoil corrections given here are exactly expressed to all orders in electron momentum and are valid for arbitrary angular structure of the photon field.

\subsection{General direction collision term for stimulated scattering}
\label{sub:General direction (stimulated) collision term}
\noindent
We now look to carry out the same procedure as in section \ref{sub:collision_term_gen_dir}, this time for the stimulated terms. 
Starting from Eq.~\eqref{app:Insert_Dopp_identity}, we find 
\begin{align}
\label{app:Insert_Dopp_identity_rotation}
\frac{\text{d}n(\nu, \vgh)}{\text{d}\tau}\Bigg|^{n^2}_{\vbh}
&\approx 4\gamma \omega \sum_{\ell m} Y_{\ell m}(\vgh) \sum_{\tilde{m}} 
\,D^{\ell}_{m \tilde{m}}(\vbh)\sum_{\tilde{\ell}}\Bigg[\delta_{\ell \tilde{\ell}}-2\beta\,\Bigg(C^{\tilde{m}}_{\ell+1}\delta_{\ell+1, \tilde{\ell}}+C^{\tilde{m}}_{\ell}\delta_{\ell-1, \tilde{\ell}}\Bigg)
\\
\nonumber
&\quad
+\beta^{2}\,\Bigg(C^{\tilde{m}}_{\ell+2}C^{\tilde{m}}_{\ell+1}\delta_{\ell+2,\tilde{\ell}}
+\Big\{(C^{\tilde{m}}_{\ell+1})^{2}+(C^{\tilde{m}}_{\ell})^{2}\Big\}\delta_{\ell\tilde{\ell}}+C^{\tilde{m}}_{\ell}C^{\tilde{m}}_{\ell-1}\delta_{\ell-2,\tilde{\ell}}\Bigg)\Bigg]
\sum_{m''}\,[D^{\tilde{\ell}}_{m'' \tilde{m}}(\vbh)]^* \mathcal{N}_{\tilde{\ell} m''}(\nu)
\end{align}
for a general boost direction. In our application, we again have to integrate over all directions of $\vbh$. With Eq.~\eqref{app:dir_avg_wigner_d}, this then yields
\begin{align}
\label{app:Insert_Dopp_identity_rotation_averaged_final}
\left<\frac{\text{d}n(\nu, \vgh)}{\text{d}\tau}\Bigg|^{n^2}_{\vbh}\,\right>
&\approx 
4\gamma \omega \sum_{\ell m} \frac{Y_{\ell m}(\vgh)}{2\ell+1} \sum_{\tilde{m}=-\ell}^\ell 
\Bigg[1+\beta^{2}\,\Big\{(C^{\tilde{m}}_{\ell+1})^{2}+(C^{\tilde{m}}_{\ell})^{2}\Big\}\Bigg]
\mathcal{N}_{\ell m}(\nu)
\nonumber\\
&= 4\gamma\omega \left(1+\frac{\beta^2}{3}\right) \sum_{\ell m} Y_{\ell m}(\vgh)\,\mathcal{N}_{\ell m}(\nu)
\nonumber\\
&\equiv 2\gamma\omega \left(1+\frac{\beta^2}{3}\right) n(\nu, \vgh)\left(2-\hat{\mathcal{O}}_{\omega}\right)\sum_{
\ell=0}^{3}c_\ell \,n_\ell(\nu, \vgh), 
\end{align}
where we used $\mathcal{N}_{\ell m}(\nu)$ given by Eq.~\eqref{app:ntilde}.
This is the boost direction averaged (stimulated) collision term contribution at {\it all} orders in $p$ and first order in electron recoil.

\section{The Kompaneets equation}
\label{Kompaneets Equation}
\noindent
In this section, we show explicitly how the Kompaneets equation emerges as a limiting case of the boost operator formalism, which itself is exact to all orders in electron momentum, $p$. Hence, we are able to clarify the physical meaning of the operator contributions. The only work we have to do is to consider the scattering process in the electron's rest frame and the rest is obtained with operator identities. Even if in the previous sections we have kept things general to all orders in $p$ and first order in the electron recoil, here we will first focus on the fastest way to obtain the desired result, but using our full knowledge of the boost operator. We will then also confirm that this is indeed what we can deduce from the more general expressions we gave above.

\subsection{The shortest path to the Kompaneets equation}
\label{sub: Kompaneets short cut}
\noindent
To obtain the Kompaneets equation using the boost operator approach in the most direct manner, we start by considering the collision term in the rest frame of the electron, but directly asking which terms remain at first order in the electron temperature. For the cross section, we have the Thomson terms, which are zeroth order in the temperature, and the recoil terms, which are already first order in the temperature once replacing $\omega_0 = \The x$, where $\The =k \Te/\me c^2\ll 1$ is the dimensionless electron temperature. This means that the recoil terms simply carry over directly from the electron rest frame, without any further ado, while for the Thomson terms we need to consider boosting effects. 

For the classical Kompaneets equation, the crucial first step is that we assume that the incoming photon distribution is isotropic in the lab frame. The Thomson part of the cross section couples to the monopole and the quadrupole anisotropy. However, boosting effects only cause leakages by $\Delta \ell=1$ for each order in $p$. That means, coupling our lab frame monopole to the quadrupole anisotropy in the electron rest frame introduces terms of second order in $p$. The same is true for the reverse coupling back to the monopole in the lab frame, meaning that any scattering of motion-induced anisotropies only enter the problem at higher orders in the temperature. In simpler words, only the Doppler operator ${}^{-1}\hat{\mathcal{D}}^0_{000}\approx 
1+\hat{\mathcal{D}}_{\nu}\frac{p^2}{3}$ contributes at first order in the electron temperature, where here we have introduced the diffusion operator $\oDnu=\oOnu^2 - 3\oOnu$. In summary, this means that for the angular dependence of the scattering cross section in the electron rest frame, we can simply write
\bealf{
\label{app:Cross_section_shortcut}
\frac{\text{d}\sigma(\omega_0, \omega, \mu_{\rm{sc}})}{\text{d}\Omega} \approx \frac{1-2\omega_0}{4\pi}
\equiv \frac{\text{d}\sigma(-\omega_0, \tilde{\omega}, \mu_{\rm{sc}})}{\text{d}\Omega},
}
without any danger of missing anything at order $\The$, as motion-induced photon anisotropies enter at higher order. This also means that in the evaluation of the collision term, we can use 
\bealf{
\label{app:photon_occ_expansion_omega_K}
n(\tilde{\omega},\vgh) \approx n(\omega_0,\vgh) - \omega_0\,\hat{\mathcal{O}}_{\omega_0}n_0(\omega_0), \qquad 
n(\omega,\vgh) \approx n(\omega_0,\vgh) + \omega_0\,\hat{\mathcal{O}}_{\omega_0}n_0(\omega_0),
}
dropping any angular dependence in the terms $\propto \omega_0$, again for similar reasons.

With these insights, from Eq.~\eqref{app:col_term_int}, the required collision term in the electron rest frame is then given by
\begin{align}
\label{app:col_term_int_K}
\frac{\text{d}n(\omega_0, \vgh_0)}{\text{d}\tau}  
&\approx 
\int 
\frac{\text{d}\vgh}{4\pi}\,\Bigg\{
n(\omega_0,\vgh)[1+n(\omega_0,\vgh_0)]-n(\omega_0,\vgh_0)[1+n(\omega_0,\vgh)]\Bigg\}
\nonumber \\
&\qquad -\omega_0 
\int 
\frac{\text{d}\vgh}{4\pi}\,\Bigg\{
[1+n(\omega_0,\vgh_0)] \hat{\mathcal{O}}_{\omega_0}n_0(\omega_0)+n(\omega_0,\vgh_0)\,\hat{\mathcal{O}}_{\omega_0}n_0(\omega_0)\Bigg\}
\nonumber\\
&\qquad \qquad +2 \omega_0 \int 
\frac{\text{d}\vgh}{4\pi}\,\Bigg\{
n(\omega_0,\vgh)[1+n(\omega_0,\vgh_0)]+n(\omega_0,\vgh_0)[1+n(\omega_0,\vgh)]\Bigg\}
\nonumber\\
&=
\int 
\frac{\text{d}\vgh}{4\pi}\,\Bigg\{
n(\omega_0,\vgh)-n(\omega_0,\vgh_0)\Bigg\}-\omega_0 [1+2n(\omega_0,\vgh_0)] \, \hat{\mathcal{O}}_{\omega_0}n_0(\omega_0)
\nonumber \\
&\qquad +2 \omega_0 \Bigg\{
n_0(\omega_0)[1+n(\omega_0,\vgh_0)]+n(\omega_0,\vgh_0)[1+n_0(\omega_0)]\Bigg\}
\nonumber\\
&\approx
n_0-n-\omega_0 [1+2n_0] \, \hat{\mathcal{O}}_{\omega_0}n_0 +4 \omega_0 n_0[1+n_0],
\end{align}
where in the last line we suppressed the arguments $\omega_0$ and $\vgh_0$. We now perform the transformations in and out of the scattering frame starting with $n_0(\omega)$ in the lab frame. This then directly yields
\begin{align}
\label{app:col_term_int_K_lab}
\frac{\text{d}n_0}{\text{d}\tau}  
&\approx
\Bigg\{{}^{-1}\hat{\mathcal{D}}^0_{000}- 1\Bigg\} \, n_0-\omega [1+2n_0] \, \hat{\mathcal{O}}_{\omega}n_0 +4 \omega n_0[1+n_0],
\end{align}
using the boost operator rules. Thermally averaging, we finally have
\begin{align}
\label{app:col_term_int_K_lab_II}
\frac{\text{d}n_0}{\text{d}\tau}  
&\approx
\The \hat{\mathcal{D}}_{\nu} n_0 +4 \omega n_0[1+n_0]+\omega^2 [1+2n_0] \, \partial_{\omega}n_0 =  \frac{\The}{\omega^2}\partial_\omega \omega^4 \partial_\omega n_0 + \frac{1}{\omega^2}\partial_\omega \Bigg\{\omega^4 n_0[1+n_0]\Bigg\},
\end{align}
where we have used $\left<p^2\right>=3\The$ and ${}^{-1}\hat{\mathcal{D}}^0_{000}- 1\approx \hat{\mathcal{D}}_{\nu}\frac{p^2}{3}$. After replacing $\omega\rightarrow \The x$, this then directly recovers the Kompaneets equation, Eq.~\eqref{Kompaneets eq}, as required. Our derivation demonstrates that the Kompaneets equation requires nothing more than the scattering of the monopole radiation field, provided that Lorentz boosts due to transformations into and out of the moving frame are taken into account. The recoil terms can be obtained by considering resting electrons, while the Doppler effects appear due to Thomson scattering. Knowledge of the properties of the boost operator and Doppler operators -- the key 'special functions' of the problem -- were enough to arrive at the final expression in a few steps once the use of the boost operator was understood.

\subsection{Kompaneets equation without stimulated scattering}
\label{sub: Kompaneets Equation - Recoil}
\noindent
We now briefly demonstrate that the more general expressions given in section~\ref{sec:Lab frame col term} also yield the Kompaneets equation at lowest order in the temperature. Assuming that the lab frame photon distribution function is isotropic, $n_{0 0}(\nu)$, in Eq. (\ref{app:fully_simplified}), we set $\ell''=0$ which also enforces $m=0$. We must also average our collision term over all boost directions, $\vbh$. Since we are considering scattering of an isotropic photon distribution only, this is equivalent to integrating with respect to all incoming photon directions, $\vgh$, which eliminates all but the monopole ($\ell=0$) terms.\footnote{This comes intuitively since any multipole other than the monopole would average to zero across the entire sky. Equivalently, the monopole itself is defined as the average of the distribution across the entire sky with higher order multipoles accounting for fluctuations.} This is the slight simplification alluded to in Section \ref{sub:collision_term_gen_dir}. The resulting collision term is given by
\bealf{
\label{app:monopole_collision}
\frac{\text{d}n_{00}(\nu)}{\text{d}\tau} &= \Bigg\{{}^{-1}\hat{\mathcal{D}}^0_{000}+\frac{1}{10}{}^{-1}\hat{\mathcal{D}}^0_{020} - 1 
\\ \nonumber
&\qquad+2\omega
\Bigg[
\left(1-\frac{1}{2}\oOnu\right)\left(
{}^{-2}\hat{\mathcal{D}}^0_{000}-\frac{2}{5}{}^{-2}\hat{\mathcal{D}}^0_{010}+\frac{1}{10}{}^{-2}\hat{\mathcal{D}}^0_{020}-\frac{3}{70}{}^{-2}\hat{\mathcal{D}}^0_{030}\right)
+\gamma\left(1+\frac{\beta^2}{3}\right)\Bigg]\Bigg\}n_{00},
}
having collapsed each of the sums and suppressed the arguments. This expression is still valid at {\it all} orders in $p$ and also directly follows from Eq.~\eqref{app:full_dir_avgd_col_term} for $\ell=m=0$ only.
Since $\The$ goes as electron momentum squared, we are interested only in those terms up to and including second order in $p$. It is therefore at this point that we throw away our all orders expansion and evaluate each of the Doppler operators up to order $p^2$. Using \verb|Mathematica| to obtain expansions of these operators, we then have 
\bsub
\bealf{
{}^{-1}\hat{\mathcal{D}}^0_{000}&\approx 
1+\hat{\mathcal{D}}_{\nu}\frac{p^2}{3}, 
&{}^{-1}\hat{\mathcal{D}}^0_{010}&\approx -\hat{\mathcal{D}}_{\nu} \frac{p^2}{3},
&{}^{-1}\hat{\mathcal{D}}^0_{020}&\approx {}^{-1}\hat{\mathcal{D}}^0_{030}\approx 0,
\\
{}^{-2}\hat{\mathcal{D}}^0_{000}&\approx 
1+\left[\frac{5}{2}+\left(\hat{\mathcal{D}}_{\nu}-\oOnu\right)\right]\,\frac{p^2}{3}, 
&{}^{-2}\hat{\mathcal{D}}^0_{010}&\approx \left[\oOnu-\hat{\mathcal{D}}_{\nu}\right] \frac{p^2}{3},
&{}^{-2}\hat{\mathcal{D}}^0_{020}&\approx {}^{-2}\hat{\mathcal{D}}^0_{030} \approx 0.
}
\esub
Writing the expansions in this way allows us to understand the origin of our terms. Consider those operators containing zeroth order terms, ${}^{-1}\hat{\mathcal{D}}^0_{000}$ and ${}^{-2}\hat{\mathcal{D}}^0_{000}$. This identity contribution derives from the fact that at zeroth order in $p$ we essentially have $\beta=0$ and thus all kernel and boost operator elements which couple between the same multipole in each frame collapse to the identity. Hence we only get leading order contributions in $p$ from those operators which describe scattering between the same multipole. The Doppler operators with $d=-1$ derive from the Thomson limit only, and thus do not account for recoil effects. This is reflected in the fact they can be expressed purely in terms of the diffusion operator.\footnote{Here, diffusion refers to the fact that despite having no energy exchange in the Thomson limit, scattered photons are Doppler shifted which, when averaged over an isotropic distribution, vanishes. However, the mean squared shift does not vanish, hence the appearance of the diffusion operator at order $p^2$. Many such scatterings like this, produce the exact analogue of Brownian motion.} The $d=-2$, operators are those for first-order recoil effects and so we get a combination of both diffusion and drift. The drift arises from the frequency changes due to recoil, hence the frequency (or energy) shift generator, $\oOnu$. We will need these terms below, but for the Kompaneets equation only the momentum dependence of ${}^{-1}\hat{\mathcal{D}}^0_{000}$ enters.

Since $\omega=\The x$, where $x=h\nu/k_B\Te$, we only need the operators in the bracket (those with $d=-2$) at leading order. Substituting these expressions into Eq. (\ref{app:monopole_collision}) and averaging over all electron momenta by using $\left<p^2\right>\approx 3\The$ \citep{CSpack2019} along with the operator identity \citep{chluba_spectro-spatial_2023-I}
\begin{equation}
    \oOnu^k = (-1)^k\sum_{m=1}^k \sterling{k}{m} \,x^m\partial_x^m,
\end{equation}
where the term in brackets denotes the Sterling numbers,
this then yields
\bealf{
\label{app:temp_collision_final}
\frac{\text{d}n_{00}(\nu)}{\text{d}\tau} 
&\approx 
\The\left\{\oDnu
+ 2 x \left(1-\frac{\oOnu}{2} + 1\right)\right\} n_{00}
\nonumber\\
&=\The \Bigg\{ 4 x \partial_x + x^2 \partial^2_x  + 4 x + x^2 \partial_x \Bigg\} n_{00} \equiv \frac{\The}{x^2}\partial_x x^4 \Big[ \partial_x n_{00} + n_{00}\Big].
}
This is the Kompaneets equation without stimulated scattering corrections. Note that the operator in the Thomson term is simply the diffusion operator, $\oDnu=\oOnu^2 - 3\oOnu = x^{-2} \partial_x x^4 \partial_x$, as we would expect excluding recoil. The second term derives from first-order recoil effects.

\subsection{Kompaneets equation -- stimulated terms}
\label{sub:Kompaneets Equation Stim}
\noindent
We now apply the same conditions to Eq.~\eqref{app:Insert_Dopp_identity_rotation_averaged_final}. Specifying an isotropic input spectrum means setting $\ell_1=\ell_2=m_1=m_2=0$ in Eq.~\eqref{app:ntilde}, which also enforces $\ell=m=0$ due to the properties of the Gaunt coefficient. At first order in $\The$, we need only take the expansion $\gamma(1+\beta^2/3)\approx 1$, as we already have a factor $\omega=\The x$ involved. This leaves us with
\begin{align}
\label{app:lab_stim_term_final}
\frac{\text{d}n_{0}(\nu)}{\text{d}\tau}\Bigg|_{\text{stim}}
\approx 2\The x \,\mathcal{G}^{0,0,0}_{0,0,0} \,n_{0}\left(2-\hat{\mathcal{O}}_{\nu}\right)n_{00}
= 2\The x \,n_{0}\left(2+x \partial_x\right) \frac{n_{00}}{\sqrt{4\pi}}
\equiv \frac{\The}{x^2}\partial_x\Big[  x^4 n^2_0\Big],
\end{align}
where we have used that $\mathcal{G}^{0,0,0}_{0,0,0}=1/\sqrt{4\pi}$ and $n_0 \equiv n_{00}Y_{00}$. This is the stimulated correction to the Kompaneets equation, which together with Eq.~\eqref{app:temp_collision_final} yields the final result in Eq.~\eqref{Kompaneets eq}.

\section{Anisotropic scattering corrections}
\label{sec: Anisotropies}
\noindent
Our derivation of the Kompaneets equation is complete, but only concerned the lowest orders in $p$ and neglected anisotropies in the photon field. However, we can now consider a few generalizations to assert whether the method is robust. The first of these is to include the scattering of the anisotropies at first order in temperature and then compare our result with current literature.

\subsection{Anisotropic scattering terms without stimulated effects}
\label{sub:Anisotropic recoil terms}
\noindent
Again, we first neglect stimulated scattering terms. For this we must use the result derived in Eq.~(\ref{app:full_dir_avgd_col_term}) for the direction-averaged collision term. 
Considering the $m$-averaged Doppler operator expansions for $\ell = 0,1,2,3$ up to $\mathcal{O}(p^2)$, we find the non-zero elements
\bsub
\bealf{
{}^{-1}\hat{\mathcal{D}}_{000}&\approx 
1+\oDnu\frac{p^2}{3},
&
{}^{-1}\hat{\mathcal{D}}_{101}&\approx 
-\left(\frac{2}{3}+\frac{\oDnu}{3}\right)\frac{p^2}{3}, 
\\
{}^{-1}\hat{\mathcal{D}}_{121}&\approx 
\left(\frac{8}{3}-\frac{2\oDnu}{3}\right)\frac{p^2}{3},
&
{}^{-1}\hat{\mathcal{D}}_{222}&\approx 
1-\left(6-\oDnu\right)\frac{p^2}{3}, 
&
{}^{-1}\hat{\mathcal{D}}_{323}&\approx 
\left(\frac{12}{7}-\frac{3\oDnu}{7}\right)\frac{p^2}{3}.
}
\esub
Note that each term can be expressed purely in terms of $\oDnu$ as we only have elements which map the lab frame multipoles to themselves via an intermediate rest frame multipole. Consequently, there is no mixing of different lab frame $\ell$ modes and thus no energy drift terms ($\propto \oOnu$). Since $\omega = x \The$ is at first order in temperature, we only require the recoil term Doppler operators evaluated at zeroth order in $p$. Hence we have ${}^{-2}\hat{\mathcal{D}}_{000}\approx
{}^{-2}\hat{\mathcal{D}}_{111}\approx
{}^{-2}\hat{\mathcal{D}}_{222}\approx
{}^{-2}\hat{\mathcal{D}}_{333}\approx 
1$, with all other elements vanishing. Additionally, we can use $\gamma(1+\beta^2/3)\approx 1$ at first order in $\The$ when combined with the $2\omega$ factor. Using $\left<p^2\right>\approx3\The$, we then have
\bealf{
\label{O plus D}
\left<\frac{\text{d}n(\nu, \vgh)}{\text{d}\tau}\Bigg|_{\vbh}\,\right>
&\approx n_0(\nu, \vgh) + \frac{1}{10}n_2(\nu, \vgh) - n(\nu, \vgh) 
\nonumber\\
&\qquad + \The\Bigg(\sum_{\ell=0}^3 \Bigg\{\beta_\ell + c_\ell\Bigg[\oDnu 
+ 2x\left(1-\frac{\oOnu}{2}\right)\Bigg]\Bigg\}\,n_\ell(\nu, \vgh)+2xn(\nu, \vgh) \Bigg),
}
with $\beta_0 = 0$, $\beta_1 = c_1$, $\beta_2 = -6c_2$, $\beta_3 = -4c_3$ and, as before, $c_0 = 1$, $c_1 = -2/5$, $c_2 = 1/10$ and $c_3=-3/70$. Disregarding the first three terms as they are simply the temperature-independent Thomson terms without any energy exchange, we can rewrite the above equation, dropping the $(\nu, \vgh)$ arguments for convenience, as
\bealf{
\label{towards c19}
\left<\frac{\text{d}n(\nu, \vgh)}{\text{d}\tau}\Bigg|_{\vbh}\,\right>
&\approx \The\sum_{\ell=1}^3\beta_\ell n_\ell + \frac{\The}{x^2}\partial_xx^4\left[\partial_x \Bar{n} + \Bar{n}\right] + 2\The x\left[n - \Bar{n}\right],
}
with $\Bar{n} = \sum_{l=0}^3c_\ell n_\ell$. The first term results from motion-induced corrections to the Thomson scattering rate. The second term is clearly the usual Kompaneets combination of diffusion and drift operators yet acting on a weighted sum of multipoles, $\Bar{n}$, not the monopole alone. It represents spectral distortion evolution and is the anisotropic generalisation of the Kompaneets operator. The third term is a recoil correction which collapses to zero when only considering the monopole. It encodes the fact that recoil acts on different multipoles with different weightings. Even though recoil effects are isotropic in the electron rest frame, when we introduce an anisotropic radiation field photons coming from regions of higher intensity contribute more to recoil than photons coming from less intense regions.

\subsection{Anisotropic stimulated terms}
\label{sub:Anisotropic stimulated terms}
We now consider the anisotropic corrections to the stimulated terms. We begin with the direction-averaged general stimulated collision term, Eq.~\eqref{app:Insert_Dopp_identity_rotation_averaged_final}.
As stimulated terms arise purely from energy exchange, and are thus $\propto \omega \propto \The$, we can again set $\gamma(1+\beta^2/3)\approx 1$. This then implies
\begin{align}
\label{app:stim_anisotropic_general_The}
\left<\frac{\text{d}n(\nu, \vgh)}{\text{d}\tau}\Bigg|^{n^2}_{\vbh}\,\right>
&\approx 2 x \The \, n(\nu, \vgh)\left(2-\hat{\mathcal{O}}_{\omega}\right)\,\bar{n}(\nu, \vgh)=2 x \The n(\nu, \vgh)\left(2+x\partial_x\right)\bar{n}(\nu, \vgh),
\end{align}
which gives the anisotropic correction to the stimulated scattering terms. Combining this result with Eq. (\ref{towards c19}), we achieve the full temperature correction,
\bealf{
\label{c19}
\left<\frac{\text{d}n(\nu, \vgh)}{\text{d}\tau}\Bigg|_{\vbh}\,\right>
&\approx \The\sum_{\ell=1}^3\beta_\ell n_\ell + \frac{\The}{x^2}\partial_xx^4\left[\partial_x \Bar{n} + \Bar{n}\right] + 2\The x\left[n - \Bar{n}\right] +2\The x\,n\left(2+x\partial_x\right)\Bar{n}
\nonumber
\\
&\approx \The\sum_{\ell=1}^3\beta_\ell n_\ell + \frac{\The}{x^2}\partial_xx^4\left[\partial_x \Bar{n} + \Bar{n}\left(1+\Bar{n}\right)\right] + 2\The x\left[n - \Bar{n}\right]\left[x+\partial_xx^2\Bar{n}\right],
}
where in the second line we regrouped terms. Note we have dropped the $(\nu, \vgh)$ arguments here for convenience. This expression is in agreement with Eq. (C19) of \cite{Chluba2012} which confirms our method has successfully derived the anisotropic correction terms to the Kompaneets equation. This conclusion is encouraging as our method is proven to work not only for the limiting Kompaneets case but also when we relax the constraints and consider additional effects.

\section{Second-order temperature corrections}
\label{app:2nd order temp}
\noindent
The second test for the formalism is to compute the second-order temperature corrections to the Kompaneets equation. Here, we will only consider monopole scattering and neglect stimulated scattering terms. This will still allow us to confirm whether or not we are correct in stating that as we go to higher order in recoil, the Doppler operators appearing should have Doppler weights growing increasingly more negative, i.e., the appearance of ${}^{-2}\hat{\mathcal{D}}_{\ell
\ell' \ell}$, an assertion that we currently cannot make as the related terms did not matter at $\mathcal{O}(\The)$.\footnote{The opposite case would mean as the order in $\omega$ increases, the Doppler weight would increase term by term starting from negative one.}

\subsection{Unstimulated collision term to second order in temperature}
\label{sub:Unstimulated collision term to second order in temperature}
\noindent
Our first amendment is to compute the collision term contributions directly $\propto \omega^2 \propto \The^2 x^2$ which arise from expansions of both the cross section and occupation number to second order in $\omega$. In the electron rest frame, expanding the cross section to second order in $\omega_0$, we have 
\bealf{
\label{app:second_order_omega}
\frac{\text{d}\sigma(\omega_0, \omega, \mu_{\rm{sc}})}{\text{d}\Omega} = \frac{3}{16\pi}\left[(1+\mu_{\rm sc}^2)-2\omega_0(1-\mu_{\rm sc})(1+\mu_{\rm sc}^2)+ \omega_0^2(1-\mu_{\rm sc})^2\left(4+3\mu_{\rm sc}^2\right)\right] + \mathcal{O}(\omega_0^3).
}
For the gain term, we must also expand $n(\tilde{\omega}, \vgh)$ to the same order in $\omega_0$, which gives
\bealf{
\label{app:photon_occ_expansion_2nd_order}
n(\tilde{\omega},\vgh) \approx n(\omega_0,\vgh) - \omega_0 (1- \mu_{\rm sc}) \,\hat{\mathcal{O}}_{\omega_0}n(\omega_0, \vgh)
+\frac{1}{2}\omega_0^2(1-\mu_{\rm sc})^2(\hat{\mathcal{O}}_{\omega_0}^2-\hat{\mathcal{O}}_{\omega_0})\,n(\omega_0, \vgh).
}
We now follow through the same steps as in section \ref{sub:Rest frame recoil terms}, expressing the terms as spherical harmonic expansions and collecting the various coefficients. Introducing $n_\ell=\sum_{m=-\ell}^\ell Y_{\ell m}(\vgh_0)\,n_{\ell m}(\omega_0)$ as before, the second-order $\omega_0$ correction to the rest frame collision term is then
\bealf{
\label{app:collision_term_second_order}
\frac{\text{d}n(\omega_0, \vgh_0)}{\text{d}\tau} \Bigg|_{\omega_0^2} 
\approx \omega_0^2\left[\sum_{\ell=0}^4 \,\left(b_{\ell}+a_{\ell}\left[\hat{\mathcal{D}}_{\omega_0}-2\hat{\mathcal{O}}_{\omega_0}\right]\right)n_{\ell}-b_0 \,n\right],
}
where, once again, $\hat{\mathcal{D}}_{\omega_0}=\hat{\mathcal{O}}_{\omega_0}^2 - 3\hat{\mathcal{O}}_{\omega_0}$ and $a_0=7/10$, $a_1=-2/5$, $a_2=1/7$, $a_3=-3/70$, $a_4=1/105$, $b_0=26/5$, $b_1=-29/10$, $b_2=67/70$, $b_3=-9/35$, $b_4=2/35$. Note the appearance of the hexadecapole term as a result of the higher order in $\The$ allowing greater mixing between multipoles (in this case the monopole coupling to the hexadecapole). Applying the same boosting method as detailed in section \ref{sub:Thomson and leading order recoil terms} to transform back to the lab frame, we have
\bealf{
\label{app:collision_term_second_order_lab}
\frac{\text{d}n(\nu, \vgh)}{\text{d}\tau} \Bigg|_{\omega^2} 
\approx \omega^2\,\sum_{\ell m \ell''} Y_{\ell m}(\vgh)\left[\sum_{\ell'=0}^4 \,\left(b_{\ell'}+a_{\ell'}\left[\hat{\mathcal{D}}_{\nu}-2\hat{\mathcal{O}}_{\nu}\right]\right)\,\Dbo{-3}{\ell\ell'\ell''}{m} -b_0 \,\Dbo{-3}{\ell\ell'\ell''}{m}\right] n_{\ell'' m}(\nu),
}
where we have redefined $\omega = h\nu/m_ec^2$ to mean the frequency as measured in the lab frame and applied $\hat{\mathcal{O}}_{\omega_0} \rightarrow\oOnu$. We now have Doppler operators with $d=-3$. This arises due to the transformation $\omega'^2\rightarrow(\nu'/\nu)^2\,\omega^2$ from the rest frame frequency $\omega'$ (formerly $\omega_0$) to the lab frame frequency $\omega$. Now, specifying monopole scattering enforces $\ell''=m=0$. Equally, since we take the average over all incoming photon directions, as before, we have $\ell=0$. Applying these conditions, we get
\bealf{
\label{app:collision_term_second_order_mon}
\frac{\text{d}n_{0 0}(\nu)}{\text{d}\tau} \Bigg|_{\omega^2} 
\approx \omega^2\,\left[\sum_{\ell'=0}^4 \,\left(b_{\ell'}+a_{\ell'}\,\left[\hat{\mathcal{D}}_{\nu}-2\hat{\mathcal{O}}_{\nu}\right]\right)\,\Dbo{-3}{0\ell'0}{0}-b_0\,\Dbo{-3}{0\ell'0}{0}\right] n_{0 0}(\nu).
}
We must again consider which Doppler operators contribute to the corrections. Since $\omega^2\propto \The^2$, we need only consider the operators which give a leading order contribution in $p$, namely the $\Dbo{-3}{000}{0}$ operator which can be approximated as the identity at this order. Finally, we can collapse the $\ell'$ sum keeping only the $\ell'=0$ term to give 
\bealf{
\label{app:collision_term_second_order_mon_0}
\frac{\text{d}n_{0 0}(\nu)}{\text{d}\tau} \Bigg|_{\omega^2} 
&\approx \omega^2\,\left[\left(b_{0}+a_{0}\,\left[\hat{\mathcal{D}}_{\nu}-2\hat{\mathcal{O}}_{\nu}\right]\right)-b_0\right]\,n_{0 0}(\nu)
=\omega^2\,a_{0}\,\left[\hat{\mathcal{D}}_{\nu}-2\hat{\mathcal{O}}_{\nu}\right]n_{0 0}(\nu).
}
The second contribution we must account for arises from the Thomson and first-order recoil terms as previously we neglected contributions $\mathcal{O}(p^3)$ from the Doppler operators. Starting from Eq. (\ref{app:monopole_collision}), for the Thomson terms we require the expansions to order $p^4$ whilst for the recoil terms we need contributions only to order $p^2$, since $\omega=\The x$. The non-zero contributors are therefore
\bsub
\bealf{
{}^{-1}\hat{\mathcal{D}}^0_{000}&\approx 
1+\hat{\mathcal{D}}_{\nu}\frac{p^2}{3}+\frac{2}{3}\left(\hat{\mathcal{D}}_{\nu}^2-4\hat{\mathcal{D}}_{\nu}\right)\frac{p^4}{15},
&{}^{-2}\hat{\mathcal{D}}^0_{000}&\approx 1+\left(\frac{5}{2}+\left(\hat{\mathcal{D}}_{\nu}-\oOnu\right)\right)\,\frac{p^2}{3},
\\
{}^{-1}\hat{\mathcal{D}}^0_{020}&\approx 
\frac{1}{3}\left(\hat{\mathcal{D}}_{\nu}^2-4\hat{\mathcal{D}}_{\nu}\right)\,\frac{p^4}{15},
&
{}^{-2}\hat{\mathcal{D}}^0_{010}&\approx (\oOnu-\hat{\mathcal{D}}_{\nu})\frac{p^2}{3},
}
\esub
with all other operators giving contributions $\mathcal{O}(p^5)$. Note, again, how those operators with $d=-1$ describe Doppler diffusion alone whilst those with $d=-2$ include energy drift contributions due to the inclusion of recoil. Combining these contributions with Eq. \eqref{app:collision_term_second_order_mon_0} and averaging over the electron momenta by applying $\left<p^2\right>\approx3\The + 15\The^2/2$  and $\left<p^4\right>\approx15\The^2$ \citep{CSpack2019}, we then have
\bealf{
\label{app:monopole_collision_expansion_again}
\frac{\text{d}n(\nu)}{\text{d}\tau} &\approx \The\Bigg\{\left(\oDnu + 4x - x\,\oOnu\right)+ \The\Bigg[\frac{5}{2}\oDnu
+ \frac{7}{10}\left(\oDnu^2 - 4\oDnu\right)+ x\left(10-\frac{137}{10}\oOnu+\frac{42}{5}\,\oOnu^2-\frac{7}{5}\,\oOnu^3\right)
\nonumber
\\
&\qquad
+\frac{7}{10}x^2\left(\oDnu-2\oOnu\right)\Bigg]\Bigg\}n(\nu)
\\
\nonumber
&\approx
\frac{\The}{x^2} \partial_x x^4 \left\{n+ \partial_x n + \frac{7}{10}\The x^2 \partial_x n + \The \left[ \frac{5}{2}(n+\partial_x n)+\frac{21}{5}x\partial_x(n + \partial_x n) + \frac{7}{10} x^2 (2\partial_x^2 n + \partial_x^3 n)\right]\right\},
}
where we have redefined $n=n(\nu)$ to mean the monopole contribution only and in the second line we have expanded each of the operators. We can confirm that this is in agreement with the terms of Eq.~(44) given by \cite{Sazonov2000} when dropping stimulated scatterings. This result can be taken as a confirmation that our Doppler weight rules are correct and that for increasing order recoil terms, the Doppler operators will have Doppler weights which decrease in integer steps. 

\subsection{Partial stimulated scattering correction at second order in temperature}
We finish by also considering the partial stimulated scattering term at first order in recoil but now at second order in $\The$. The thermal average of $\gamma(1+\beta^2/3)$ yields  $\left<\gamma(1+\beta^2/3)\right>\approx\left< 1+ \frac{5}{6} p^2 \right>\approx 1+\frac{5}{2} \The$. This then implies the second order temperature correction
\begin{align}
\label{app:lab_stim_term_final_II}
\frac{\text{d}n_{0}(\nu)}{\text{d}\tau}\Bigg|_{\text{stim}}
\approx \left(1+\frac{5}{2} \The\right) \frac{\The}{x^2}\partial_x\Big[  x^4 n^2_0\Big],
\end{align}
up to first order in recoil. Comparing to \cite{Sazonov2000}, we again confirm the correctness of this contribution. 
We did not consider the second order recoil terms in this work. However, we expect the boost operator method to be applicable in a similar matter and yield the correct final result.

\section{Conclusion}
\label{Conclusion}
\noindent
In this work, we have presented an elegant derivation of the Kompaneets equation including stimulated terms using the recently developed boost operator approach (see \cite{ChlubaBO25}). A 'shortcut' derivation is provided in section \ref{sub: Kompaneets short cut}, using our full knowledge of the new formalism, emphasising the power and simplicity of the approach. We further derived exact general expressions for the Compton collision term to first order in the electron recoil ($\propto h\nu/\me c^2$) in terms of boost and Doppler operators which retain all orders in electron momentum [see Eq.~(\ref{app:fully_simplified}) and Eq.~(\ref{app:Insert_Dopp_identity})]. 
For isotropic electron distributions, we also performed the average over all directions of the electron momenta [see Eq.~\eqref{app:full_dir_avgd_col_term} and Eq.~\eqref{app:Insert_Dopp_identity_rotation_averaged_final}].
The required operators can be expanded in powers of the energy-shift generator, $\oOnu\equiv-\nu\partial_{\nu}$, using the purposefully developed \verb|Mathematica| notebook.\footnote{This can be found at \url{https://www.jb.man.ac.uk/~jchluba/Science/Mathematica/}.} 
Furthermore, we have demonstrated that the boost operator approach reproduces the known results for the anisotropic scattering corrections (see section~\ref{sec: Anisotropies}) and the second-order temperature corrections (see section~\ref{app:2nd order temp}) to the Kompaneets equation, thereby validating the approach. Higher-order corrections can in principle be computed by the same method, but are left to future work. 

The generality of the boost operator formalism suggests its applicability to finding the kinematic corrections to the Kompaneets equation as well as to photon scattering processes beyond Comptonisation. Moreover, while here we performed a series expansion in orders of the electron recoil,
a continuation of this work might explore a possible resummation of Doppler operators of decreasing Doppler weight. This may yield scattering operators exact to all orders in recoil, eliminating the need for term-by-term evaluation. However, for this, one will first have to study the symmetries and properties of the Doppler operator more carefully, as significant simplifications are expected.

Finally, we have also provided a new expression for the aberration kernel and boost operator for arbitrary boost direction in terms of the well-known kernel for a boost collinear with $\hat{\boldsymbol{z}}$ [see Eq. (\ref{app:kernel_elements_spherical})]. The related expressions should be useful for a wide range of applications that we look forward to exploring in the future.

\small 

\vspace{2mm}

\noindent
{\it Data Availability Statement}: {\tt Mathematica} files to reproduce some of the key results are available at \url{https://www.jb.man.ac.uk/~jchluba/Science/Mathematica/}.

{\small
\bibliographystyle{plain}
\bibliography{Lit-all}
}

\appendix
\section{Symmetries of the aberration kernel}
\label{app:Kernel symmetries}
\noindent
Here we label a few of the general kernel symmetries outlined by \cite{Dai2014} along with some recurrence relations useful in calculating kernel elements. Some of these are extensively used in the \verb|Mathematica| notebook\footnote{This can be found at \url{https://www.jb.man.ac.uk/~jchluba/Science/Mathematica/}}.
For full derivations of these see \cite{Dai2014}.
Beginning from the kernel definition in the coordinate system where $\vbh \parallel \boldsymbol{\hat{z}}$, 
\bealf{
\label{app:Aberration Kernel with s}
{}^{d}_{s}\mathcal{K}_{\ell' \ell}^{m}\,(\beta)
&= 
\int \text{d}^2\boldsymbol{\hat{n}}'\,
\frac{{}_{-s}Y_{\ell' m}^{*}(\theta', \phi')\,{}_{-s}Y_{\ell m}(\theta, \phi)}{\left[\gamma \, (1-\beta \, \mu')\right]^{\,d}},
}
where we have restored the spin weight and note that $\boldsymbol{\hat{n}}'=\boldsymbol{\hat{n}}'\,(\theta', \phi')$. We can imagine a transformation such that $\theta \xrightarrow{} \pi-\theta$, $\phi \xrightarrow{} \phi + \pi$ and $\beta \xrightarrow{} -\beta$. Under this transformation we also have $\theta' \xrightarrow{} \pi-\theta'$, $\phi' \xrightarrow{} \phi' + \pi$ such that the kernel remains unchanged. Since it is true that
\bealf{
\label{app:Y transformation}
{}_{s}Y_{\ell m}(\pi - \theta, \phi + \pi)
= (-1)^\ell {}_{-s}Y_{\ell m}(\theta, \phi),
}
we can see that, by inspection,
\bealf{
\label{app:Kernel relation 2}
{}^{d}_{s}\mathcal{K}_{\ell' \ell}^{m}\,(-\beta)
&= 
(-1)^{\ell +\ell'}\,{}^{d}_{-s}\mathcal{K}_{\ell' \ell}^{m}\,(\beta).
}
Thus we see a symmetry of the kernel under reversal of the boost axis. Again, starting from Eq. (\ref{app:Aberration Kernel with s}), we can use the identity ${}_{-s}Y_{\ell' -m}(\theta, \phi) = (-1)^{s+m}\,[{}_{s}Y_{\ell' m}(\theta, \phi)]^*$ on both spherical harmonic functions to write
\bealf{
\label{app:Aberration Kernel with s reversed}
{}^{d}_{s}\mathcal{K}_{\ell' \ell}^{m}\,(\beta)
&= 
\int \text{d}^2\boldsymbol{\hat{n}}\,
\frac{{}_{s}Y_{\ell' -m}^{*}(\theta', \phi')\,{}_{s}Y_{\ell -m}(\theta, \phi)}{\left[\gamma \, (1-\beta \, \mu')\right]^{\,d}}
\equiv {}^{d}_{-s}\mathcal{K}_{\ell' \ell}^{-m}\,(\beta).
}
Another symmetry derived in Appendix B of \cite{Dai2014} is 
\bealf{
\label{app:Kernel ell relation}
{}^{d}_{s}\mathcal{K}_{\ell \ell'}^{m}\,(\beta)
&= 
(-1)^{\ell+\ell'}{}^{2-d}_{-s}\mathcal{K}_{\ell' \ell}^{m}\,(\beta),
}
for which the derivation is slightly more involved. Combining the results of Eq. (\ref{app:Kernel relation 2}) and Eq. (\ref{app:Kernel ell relation}) we have
\bealf{
\label{app:Kernel beta relation}
{}^{d}_{s}\mathcal{K}_{\ell \ell'}^{m}\,(\beta)
&= 
{}^{2-d}_{s}\mathcal{K}_{\ell' \ell}^{m}\,(-\beta).
}
This underlines the symmetry of the kernel under the exchanges $\ell \xrightarrow{} \ell'$, $\ell' \xrightarrow{} \ell$ and $\beta \xrightarrow{} -\beta$. We also note two more identities which raise or lower the Doppler weights of kernels for $s=0$. They are given by
\bsub
\bealf{
\label{app:Kernel doppler relation}
{}^{d}_{}\mathcal{K}_{\ell' \ell}^{m}\,(\beta)
&= 
\gamma \,{}^{d+1}_{}\mathcal{K}_{\ell' \ell}^{m}\,(\beta) -\gamma \beta \left[ C^{m}_{\ell'+1}\,{}^{d+1}_{}\mathcal{K}_{\ell'+1 \ell}^{m}\,(\beta) + C^{m}_{\ell'}\,{}^{d+1}_{}\mathcal{K}_{\ell'-1 \ell}^{m}\,(\beta)\right],
\\
\label{app:Kernel doppler relation 2}
&= 
\gamma \,{}^{d-1}_{}\mathcal{K}_{\ell' \ell}^{m}\,(\beta) +\gamma \beta \left[ C^{m}_{\ell+1}\,{}^{d-1}_{}\mathcal{K}_{\ell' \ell+1}^{m}\,(\beta) + C^{m}_{\ell}\,{}^{d-1}_{}\mathcal{K}_{\ell' \ell-1}^{m}\,(\beta)\right],
}
\esub
where $C^{m}_{\ell} = \sqrt{(\ell^2-m^2)/(4\ell^2-1)}$ for $\ell > 0$ and $\ell \geq |m|$. The full expressions (including for non-zero $s$) are derived in \cite{Dai2014}. Recursion relations linking elements of differing $\ell$ values also exist. Given in Appendix B of \cite{Chluba2026SZ} is the expression
\bealf{
\label{eq: kernel change l}
^{d}_{}\mathcal{K}_{\ell \ell'}^{0}(-\beta)
&= 
-\sqrt{\frac{4\ell^2-1}{\ell^2}}\frac{\gamma\,^{d}_{}\mathcal{K}_{\ell-1 \ell'}^{0}(-\beta)-\,^{d-1}_{}\mathcal{K}_{\ell-1 \ell'}^{0}(-\beta)}{p}-\frac{\ell-1}{\ell}\sqrt{\frac{2\ell+1}{2\ell-3}}^{d}_{}\mathcal{K}_{\ell-2 \ell'}^{0}(-\beta),
}
which can be used to generate the all $^{d}_{}\mathcal{K}_{\ell 0}^{0}(-\beta)$ starting from the directly evaluated $^{d}_{}\mathcal{K}_{0 0}^{0}(-\beta)$. Using Eq. (\ref{app:Kernel beta relation}), the elements $^{d}_{}\mathcal{K}_{0 \ell'}^{0}(-\beta)$ can then be calculated. Finally, implementing Eq. (\ref{eq: kernel change l}) for each element generates all remaining elements $^{d}_{}\mathcal{K}_{\ell \ell'}^{0}(-\beta)$. This procedure is efficiently implemented in the \verb|Mathematica| notebook, demonstrating the ease at which exact kernel elements may be computed, avoiding complicated integrals.

\section{Gaunt integrals}
\label{app:Gaunt}
Integrals over products of three spherical harmonics can be nicely written in terms of the Gaunt coefficients
\bealf{
\label{eq:Gaunt}
\mathcal{G}^{\ell,\ell_1,\ell_2}_{m,m_1,m_2}
&=
\int \id \hat{\vek{r}}\,
Y_{\ell m}(\hat{\vek{r}})
\,Y_{\ell_1 m_1}(\hat{\vek{r}})
\,Y_{\ell_2 m_2}(\hat{\vek{r}})
\nonumber\\
&
=\sqrt{
\frac{(2\ell+1)(2\ell_1+1)(2\ell_2+1)}{4\pi}
}
\,\Bigg(
\begin{array}{ccc}
\ell  & 
\ell_1 & \ell_2\\
0 & 0 & 0
\end{array}
\Bigg)
\,\Bigg(
\begin{array}{ccc}
\ell  & \ell_1 & \ell_2\\
m & m_1 & m_2
\end{array}
\Bigg),
}
where we used the Wigner-$3J$ symbols. These enforce $m+m_1+m_2=0$, which means that $\mathcal{G}^{\ell,\ell_1,\ell_2}_{m,m_1,-m_1}$ means $m=0$. Similarly, $(-1)^m \mathcal{G}^{\ell,\ell_1,\ell_2}_{-m,m_1,m_2}$ appears when the first spherical harmonic is conjugated, as often relevant to projection integrals. We also have $[\mathcal{G}^{\ell,\ell_1,\ell_2}_{m,m_1,m_2}]^*=\mathcal{G}^{\ell,\ell_1,\ell_2}_{-m,-m_1,-m_2}$.

We will also encounter integrals over three spin-harmonics, which can be written as
\bealf{
\label{eq:Gaunt2}
{}_{s, s_1, s_2}\mathcal{G}^{\ell,\ell_1,\ell_2}_{m,m_1,m_2}
&=
\int \id \hat{\vek{r}}\,
{}_{-s}Y_{\ell m}(\hat{\vek{r}})
\,{}_{-s_1}Y_{\ell_1 m_1}(\hat{\vek{r}})
\,{}_{-s_2}Y_{\ell_2 m_2}(\hat{\vek{r}})
\nonumber\\
&
=\sqrt{
\frac{(2\ell+1)(2\ell_1+1)(2\ell_2+1)}{4\pi}
}
\,\Bigg(
\begin{array}{ccc}
\ell  & 
\ell_1 & \ell_2\\
s & s_1 & s_2
\end{array}
\Bigg)
\,\Bigg(
\begin{array}{ccc}
\ell  & \ell_1 & \ell_2\\
m & m_1 & m_2
\end{array}
\Bigg),
}
which requires $s+s_1+s_2=0$.
%

\end{document}